\documentclass{article}
\usepackage[T1]{fontenc}
\usepackage{graphicx}
\usepackage{subcaption}
\usepackage{amsmath}
\usepackage{amsthm}
\newtheorem{prpstn}{Proposition}
\numberwithin{equation}{section}
\usepackage{amsfonts}
\usepackage{txfonts}
\usepackage{dsfont}
\usepackage{latexsym}
\usepackage[empty]{fullpage}
\usepackage{marvosym}
\usepackage[usenames,dvipsnames]{color}
\usepackage{verbatim}
\usepackage{enumitem}
\usepackage{fancyhdr}
\usepackage{tabularx}
\usepackage{microtype}
\usepackage{fontawesome5}
\usepackage{pifont}
\usepackage[top=2.5cm, bottom=3.5cm, left=2.5cm, right=2cm]{geometry}
\usepackage[colorlinks,urlcolor=blue,breaklinks=true,linkcolor=blue,citecolor=blue]{hyperref}
\usepackage[T1]{fontenc}
\usepackage{xcolor} 
\usepackage{marginnote}
\usepackage{chngcntr}
\usepackage{algorithm, algpseudocode}
\usepackage{multirow}

\newcommand\bs[1]{\boldsymbol{#1}}

\newcommand{\tr}{\text{trace}} 
\newcommand{\E}{\mathbb{E}}
\newcommand{\p}{\mathbb{P}}
\newcommand{\GP}{\text{GP}}
\newcommand{\R}{\mathbb{R}}
\newcommand{\argminE}{\mathop{\mathrm{argmin}}} 
\newcommand{\ValentinCorrections}[1]{#1}
\newcommand{\SecondRoundValentinCorrections}[1]{#1}
\newcommand{\BreazBristol}{School of Civil, Aerospace and Design Engineering, University of Bristol, United Kingdom; e-mail: valentin.breaz@bristol.ac.uk}
\newcommand{\RichardNotts}{School of Mathematical Sciences, University of Nottingham, United Kingdom; e-mail: r.d.wilkinson@nottingham.ac.uk}
\newcommand{\InriaAIR}{Univ. Grenoble Alpes, Inria, CNRS, Grenoble INP*, LJK, 38000 Grenoble, France; e-mail: olivier.zahm@inria.fr}
\newcommand{\IFPEN}{IFP \'Energies Nouvelles, France; e-mail: miguel.munoz-zuniga@ifpen.fr}
\title{\vspace{-2cm}A new gradient-free active subspace estimation method with application to rare event probability estimation}
\date{\vspace{-3ex}November 2026}
\begin{document}
\maketitle

{\centering
  \large\textbf{Valentin Breaz\footnotemark[1], Miguel Munoz Zuniga\footnotemark[2], Olivier Zahm\footnotemark[3], Richard Wilkinson\footnotemark[4]}
  \par}
\footnotetext[1]{\BreazBristol}
\footnotetext[2]{\IFPEN}
\footnotetext[3]{\InriaAIR}
\footnotetext[4]{\RichardNotts}
\tableofcontents
\begin{abstract}
In order to reduce the high cost of estimating the probability of a rare event involving a very large number of random parameters, we propose a new strategy for dimension reduction coupled with a surrogate model for the expensive part of the algorithm.
To this end, we extend the Ordinary Kriging Active Subspace (OK-AS) method \cite{Palar2018} into a sequential version.
OK-AS constructs a high-dimensional Gaussian Process (GP) model from which the dimension reduction matrix determining the active subspace \cite{Constantine2014, Zahm2020} can be estimated using the derivatives of the GP posterior mean. 
The approach proposed here consists of iteratively re-estimating the active subspace from a GP trained in a rotated coordinate system until the active subspace stabilises. 
Our sequential approach, no more costly in simulations than the initial version, allows for a reduction in prediction error and a better approximation of the active subspace on a benchmark of test problems. 
For comparison, a large number of state-of-the-art dimension reduction methods are also considered. Furthermore, we integrate our algorithm into an efficient pre-existing approach for estimating the probability of a rare event \cite{Uribe2021cross}. 
This approach is based on learning the active subspace associated with the random event whose probability is to be estimated. The sequential learning of an importance sampling density is necessary and corresponds to the expensive part of this strategy \cite{Uribe2021cross}. 
To circumvent this, we integrate our sequential OK-AS version into the estimation process of the importance sampling density. We compare our method against two efficient strategies: Sequential Importance Sampling (SIS) \cite{Papaioannou2016sequential} 
and Active Learning Sequential Subspace Importance Sampling (ASSIS) \cite{Ehre2022}. The numerical results, on a set of test and industrial problems, indicate that our method allows for reducing the number of simulations required to obtain a precise estimate of the rare
event probability.     
\end{abstract}

\section{Introduction}\label{sec:intro}
In engineering, the state of a system is often characterized by a function $Q(\mathbf{x}):\R^d\to\R$. The failure set, denoted $\mathcal{L}_f$, is the region of the input space where the state function does not meet a certain threshold $\beta$:
\begin{equation}\label{eq:excursion_set}
\mathcal{L}_f:=\big\{\mathbf{x}\in\R^d : f(\mathbf{x}):=Q(\mathbf{x})-\beta\leq 0\big\}.     
\end{equation}
Thus, $\mathcal{L}$ can be seen as an operator that associates a set to a function $f$. The system inputs are modelled as a random vector $\mathbf{X}$. Our objective is to compute the failure probability $P_f=\p_\mathbf{X}[f(\mathbf{X})\leq 0]$ in high dimensions (i.e., $d \gg 1$), where $\{f(\mathbf{X})\leq 0\}$ is often a rare event. We can use an indicator function to express this probability as:
\begin{align}\label{eq:MC_rare_events}
P_f=\E_\mathbf{X}[\mathbb{I}(f(\mathbf{X})\leq 0)].
\end{align}
\par There exists a large number of methods for estimating $P_f$ (\ref{eq:MC_rare_events}). If the simulator $f$ is linear or weakly nonlinear, the most popular approach is the family of methods known as first-order reliability
(FORM) \cite{der_kiureghian_multiple_1998, Wang2017_form}. FORM methods reformulate (\ref{eq:MC_rare_events}) as a constrained optimization problem. If the simulator $f$ is nonlinear, many other approaches are available, such as standard Monte Carlo (MC) \cite{liu2001monte} 
and second-order reliability (SORM) \cite{ditlevsen_structural_1996}. However, standard MC sampling of $\mathbf{X}$ can be too expensive, especially for complex simulators, or for problems such as rare event estimation (e.g., with $P_f\leq10^{-3}$) where a large number of 
samples are required. Variance reduction methods such as Subset Simulation
\cite{au_estimation_2001}, Importance Sampling using FORM initialization \cite{schueller_critical_1987}, Sequential Importance Sampling \cite{Papaioannou2016sequential}, and asymptotic sampling \cite{BUCHER2009504} have demonstrated improved cost-efficiency with respect to 
standard MC in various applications. Nonetheless, these methods can sometimes still be too expensive. If a surrogate approximation of $f$ exists (e.g., using Gaussian Processes (GP) or Polynomial Chaos Expansions (PCE)) that is sufficiently accurate and affordable to construct, 
the rare event probability estimation cost can be greatly reduced \cite{moustapha2022active}. However, high-dimensional problems can pose difficulties for surrogate methods regarding training costs, and for Markov Chain Monte Carlo (MCMC) samplers as it can lead 
to slow convergence. MCMC samples are used in popular variance reduction strategies such as Subset Simulation \cite{au_estimation_2001} and Sequential Importance Sampling \cite{Papaioannou2016sequential}. \ValentinCorrections{While specialized MCMC methods for high-dimensional rare event problems have been developed \cite{Papaioannou2015}, they can still require a very high computational cost in the context of expensive simulators \cite{MMZ2021}.} 
As a result, dimension reduction methods have been used in some rare event 
probability estimation problems. Good performance has been demonstrated by Active Subspace (AS) \cite{Jiang2017}, Partial Least Squares (PLS) \cite{zuhal2021dimensionality}, Kernel Dimension Reduction (KDR) \cite{MMZ2021}, one-dimensional projections \cite{ELMASRI2021107991},
and Sliced Inverse Regression (SIR) \cite{pan2017sliced}, to name only a few dimension reduction methods.
\par Another point of view consists in seeing $P_f$ as the volume of the failure set $\mathcal{L}_f$ in (\ref{eq:excursion_set}) with respect to the measure $\nu_\mathbf{X}$ associated with the random vector $\mathbf{X}$:
\begin{align}\label{eq:MC_rare_events_volume}
P_f=[\mathcal{V}\circ \mathcal{L}]_f
\end{align}
where for any compact set $A$, $\mathcal{V}(A)=\nu_\mathbf{X}(A)=\E_\mathbf{X}(\mathbb{I}_A)$. 
In terms of solving strategies, the two formulations (\ref{eq:MC_rare_events}) and (\ref{eq:MC_rare_events_volume}) are not equivalent. In particular, methods for adaptively acquiring simulations differ depending on the objective (\ref{eq:MC_rare_events}) or 
(\ref{eq:MC_rare_events_volume}). In the first formulation (\ref{eq:MC_rare_events}), the problem is solved by estimating an expectation of a potentially non-differentiable function \cite{madsen_methods_1986, ditlevsen_structural_1996, moustapha2022active}. 
In the second formulation (\ref{eq:MC_rare_events_volume}), the objective is decomposed into the estimation of an excursion set $\mathcal{L}_f$, whose volume is then calculated or estimated \cite{Bect2012, azzimonti_adaptive_2021}.
\par In this work, we treat the reliability problem as an integration problem in the sense of (\ref{eq:MC_rare_events}). In Section \ref{sec:ce_seqsaas}, we propose a sequential strategy for the estimation of the probability of a rare event, 
when the function $f$ is computationally expensive, and when $\mathbf{X}$ is a high-dimensional random vector. These constraints greatly complicate any statistical approach necessary for its estimation. For this reason, in Section 
\ref{sec:seqokas_introduced} we propose a new Gaussian Process (GP) based strategy to counter the problem of expensive simulation. Our strategy involves a new gradient-free dimension reduction method, whose advantages are demonstrated numerically in Section
\ref{sec:numerical_results}. Another contribution presented in Section \ref{sec:numerical_results} is a method comparison between a large number of approaches for dimension reduction and high-dimensional GP regression on a benchmark of test problems. Finally, 
Section \ref{sec:method_comparisonzz} shows that our approach is competitive for various non-linear, high-dimensional reliability problems.

\section{Dimension reduction for a high-dimensional Gaussian Process}\label{sec:dim_red_gp}
We will denote by $F_n$ the Gaussian process, conditioned on $n$ data points, modeling $f(\mathbf{x}):\mathbb{R}^d\to\mathbb{R}$.
A major drawback of GP modeling is the curse of dimensionality that occurs as the dimension of ${d}=\operatorname{dim}\mathbf{x}$ increases. A prohibitive number of samples $n$ may be required to build an accurate model \cite{Stone1980}. In practice, a common heuristic is to use $n\leq d^2$ \cite{Binois2022survey}.
Fortunately, in practice many simulators $f$ are effectively low-dimensional, and can be characterized by a "ridge" type approximation:
$$ f(\mathbf{x})\approx f_r(\mathbf{W}^T\mathbf{x}), \quad \text{where } \mathbf{W}\in\mathbb{R}^{d\times r} \text{ and } r\ll d.$$
\par There are different methods for finding a ridge structure $\mathbf{W}$, and thus construct a low-dimensional GP surrogate given $n$ training points $\{\mathbf{W}^T\mathbf{x_i}, f(\mathbf{x_i})\}_{i=1}^n$, e.g., Kernel Dimension Reduction \cite{kdr_fukumizu_et_al}, Embedding Learning \cite{NIPS1998_ed422773}, Partial Least Squares \cite{bouhlel2016improving}, Active Subspaces \cite{Constantine2014}.

\subsection{Summary review of existing methodologies}\label{sec:dim_red_summary}
\noindent\textbf{Kernel Dimension Reduction (KDR).} Sufficient Dimension Reduction (SDR) methods involve searching for $\mathbf{W_\star}\in\mathbb{R}^{d\times r}$ such that conditional independence is achieved: $f(\mathbf{X})\perp \mathbf{X}|\mathbf{W_\star}^T\mathbf{X}$. 
This is equivalent to saying there exists a measurable function $g$ such that $f(\mathbf{X})=g(\mathbf{W_\star}^T\mathbf{X})$ \cite{glaws2020inverse}. The approach known as Kernel Dimension Reduction (KDR) was introduced in \cite{kdr_fukumizu_et_al} as an SDR method.
 KDR remains a principled approach even when conditional independence does not hold exactly, but only in the sense of a ridge approximation: $f(\mathbf{X})\approx g(\mathbf{W_\star}^T\mathbf{X})$. The KDR estimator uses kernel functions 
 $k_X(\mathbf{W}^T\mathbf{x}, \mathbf{W}^T\mathbf{x'})$ and $k(f(\mathbf{x}), f(\mathbf{x'}))$ (e.g., Radial Basis Function (RBF) kernels) defined on the reduced input space $\mathbb{R}^r$ ($\mathbf{W}\in\mathbb{R}^{d\times r}$) and the output space $\mathbb{R}$, respectively:
\begin{equation}\label{eq:kdr_fukumizu}
\mathbf{\hat{W}_{KDR}}=\argminE_{\mathbf{W}:\mathbf{W}^T\mathbf{W}=\mathbf{I_r}} \mathrm{Tr}[\mathbf{G_Y}(\mathbf{G_X^W}+n\varepsilon_n\mathbf{I_n})^{-1}],
\end{equation}
where $\mathcal{X}=\{\mathbf{x_i}, f(\mathbf{x_i})\}_{i=1}^n$ is the training set, $\mathbf{G_Y}:=(\mathbf{I_n}-\frac{1}{n}\mathbf{1}\mathbf{1}^T)\mathbf{K_Y}(\mathbf{I_n}-\frac{1}{n}\mathbf{1}\mathbf{1}^T)$ is the centred Gram matrix 
$\mathbf{K_Y}:=(k(f(\mathbf{x_i}), f(\mathbf{x_j})))_{i, j=1}^n$ (similarly for $\mathbf{G^W_X}$), and $\varepsilon_n$ provides regularization for matrix inversion. Steepest descent optimization with line search
\cite{kdr_fukumizu_et_al} can be used to solve the non-convex optimization problem (\ref{eq:kdr_fukumizu}). The KDR estimator $\mathbf{\hat{W}_{KDR}}\in\mathbb{R
}^{d\times r}$ and the resulting training set $\{\mathbf{\hat{W}}^T_{\mathbf{KDR}}\mathbf{x_i}, f(\mathbf{x_i})\}_{i=1}^n$ were used in order to construct a low-dimensional GP surrogate for $f(\mathbf{x})$:
\begin{equation}\label{eq:gp_kdr}
g^{KDR}_{GP}(\mathbf{\hat{W}}^T_{\mathbf{KDR}}\mathbf{x})\sim\GP(m_{KDR}(\mathbf{\hat{W}}_{\mathbf{KDR}}^T\mathbf{x}), k_{KDR}(\mathbf{\hat{W}}_{\mathbf{KDR}}^T\mathbf{x}, \mathbf{\hat{W}}_{\mathbf{KDR}}^T\mathbf{x'})).
\end{equation}
See, for example, \cite{MMZ2021} where an expensive function derived from a wind turbine simulator was approximated. One downside of KDR is that large training times are reported in the literature when $d$ is large. In this regard, a faster SDR approach known as gradient 
Kernel Dimension Reduction (gKDR) was introduced in  \cite{Fukumizu2014}. The gKDR estimator $\mathbf{\hat{W}_{gKDR}}\in\mathbb{R}^{d\times r}$ consists of the dominant $r$ eigenvectors of the following matrix:
\begin{equation}\label{eq:gkdr_fukumizu}
\mathbf{\hat{M}_{gKDR}}=\frac{1}{n}\sum_{i=1}^n \nabla k_X(\mathbf{x_i})^T(\mathbf{K_X}+n\varepsilon_n I)^{-1}\mathbf{K_Y} (\mathbf{K_X}+n\varepsilon_n I)^{-1}\nabla k_X(\mathbf{x_i}),    
\end{equation}
where $\mathbf{K_X}=(k_X(\mathbf{x_i}, \mathbf{x_j}))_{i, j=1}^n$ is the Gram matrix for a kernel $k_X$ (e.g., RBF kernel), $\mathbf{K_Y}:=(k(f(\mathbf{x_i}), f(\mathbf{x_j})))_{i, j=1}^n$ is the Gram matrix for a kernel $k(f(\mathbf{x}), f(\mathbf{x'}))$,
and $\nabla k_X(\mathbf{x}):=\big(\frac{\partial k_X(\mathbf{x_1}, \mathbf{x})}{\partial \mathbf{x}},\dots,\frac{\partial k_X(\mathbf{x_n}, \mathbf{x})}{\partial \mathbf{x}}\big)$ is the Jacobian matrix.\\
\noindent\textbf{Embedding Learning (EL).} An alternative approach is to replace $\mathbf{\hat{W}_{KDR}}$ in (\ref{eq:gp_kdr}) with a learned embedding $\mathbf{\hat{W}_{EL}}\in\mathbb{R
}^{d\times r}$ that maximizes the marginal log-likelihood for $g_{GP}(\mathbf{{W}}^T_{\mathbf{GP}}\mathbf{x})\sim\GP(m({\mathbf{W}}_{\mathbf{GP}}^T\mathbf{x}), k({\mathbf{W}}_{\mathbf{GP}}^T\mathbf{x}, {\mathbf{W}}_{\mathbf{GP}}^T\mathbf{x'}))$:
$$    \log p(\mathbf{y}=(f(\mathbf{x_i}))_{i=1}^n|\mathbf{{W}}_{\mathbf{GP}})=-\frac{1}{2}(\mathbf{y}-(m({\mathbf{W}}_{\mathbf{GP}}^T\mathbf{x_i}))_{i=1}^n)^T\mathbf{K_X}^{-1}(\mathbf{y}-(m({\mathbf{W}}_{\mathbf{GP}}^T\mathbf{x_i}))_{i=1}^n)-\frac{1}{2}\log\det(\mathbf{K_X})-\frac{n}{2}\log 2\pi$$
with respect to $\mathbf{W_\GP}$, where $\mathbf{K_X}=(k({\mathbf{W}}_{\mathbf{GP}}^T\mathbf{x_i}, {\mathbf{W}}_{\mathbf{GP}}^T\mathbf{x_j}))_{i, j=1}^n$ is the Gram matrix. 
In other words, the projection matrix $\mathbf{W_\GP}\in\mathbb{R}^{d\times r}$ can be seen as an additional set of GP hyperparameters to be optimized in training. This method is known as Embedding Learning (EL) \cite{NIPS1998_ed422773}.\\
\noindent\textbf{Partial Least Squares (PLS).} The classical dimension reduction method Partial Least Squares (PLS) \cite{wold1966estimation} was introduced for GP surrogates by \cite{bouhlel2016improving}. Using the training set $\mathcal{X}=\{\mathbf{x_i}, f(\mathbf{x_i})\}_{i=1}^n$ generated 
from $n$ independent sample of $\mathbf{X}$, the PLS matrix $\mathbf{\hat{W}_{PLS}}\in\mathbb{R}^{d\times r}$ is constructed iteratively using a sequence of linear regressions. On one hand, PLS is less flexible compared to KDR and EL, since it is not coupled with any non-linear regression
method. On the other hand, PLS is less prone to bad local minima, since both KDR and EL require solving non-convex, high-dimensional optimization problems.
The resulting low-dimensional GP surrogate trained on $\{\mathbf{\hat{W}}^T_{\mathbf{PLS}}\mathbf{x_i}, f(\mathbf{x_i})\}_{i=1}^n$ has been used in different areas, e.g., rare event probability estimation \cite{sehic7}. The predictive performance of GP-PLS has been further improved in \cite{bouhlel_improved_2016}, where a method known as GP-PLSK has been developed.
The idea of GP-PLSK is to train a high-dimensional GP using PLS-informed hyperparameter initialisation. However, the GP-PLSK training time is often significantly longer compared to GP-PLS.\\
\noindent\textbf{Other linear dimension reduction approaches.}
The problem of dimension reduction can also be formulated as finding a projection matrix $\mathbf{W}\in\R^{d\times r}$ that minimizes the following approximation error:
\begin{equation}\label{eq:l2_error_ridge}
\mathcal{E}(\mathbf{W}):= \min_{f_r:\R^r\to\R} \left(\int|f(\mathbf{x})-f_r(\mathbf{W}^T\mathbf{x})|^2q(\mathbf{x})d\mathbf{x}\right)^{1/2},
\end{equation}
where $q(\mathbf{x})$ is the density of the uncertain input vector $\mathbf{X}$. The optimal function $f_r$ is the conditional expectation $f_r(\mathbf{x}_r) = \mathbb{E}_\mathbf{X}[f(\mathbf{X})|\mathbf{W}^T \mathbf{X}=\mathbf{x}_r]$. Two options exist: 
minimize $\mathcal{E}(\mathbf{W})$ directly by explicitly estimating $f_r$, which can be expensive, or minimize an upper bound of $\mathcal{E}(\mathbf{W})$, which avoids this expensive calculation.\\
\noindent\textbf{Active Subspace (AS).} The Active Subspace (AS) method \cite{Constantine2014} proposes to define $\mathbf{W}\in\R^{d\times r}$ as the matrix of the $r$ dominant eigenvectors of the gradient covariance matrix:
\begin{equation}\label{eq:as_with_grad}
 \mathbf{H}=\int \nabla f(\mathbf{x})\nabla f(\mathbf{x})^T q(\mathbf{x})d\mathbf{x}.
\end{equation}
This approach minimizes an upper bound of the error $\mathcal{E}(\mathbf{W})$ (\ref{eq:l2_error_ridge}) \cite{Zahm2020}. The advantage of AS is its simplicity, but its major drawback is the need for gradients of $f$, which are often unavailable for complex simulators. However, 
AS has been used as a dimension reduction tool for GPs in areas such as geophysics \cite{ma2022computer}, engineering \cite{parente2020active}, COVID-19 models \cite{Wycoff2021}, and Ebola spread models \cite{parente2020active}.\\
\noindent\textbf{Gradient-free Active Subspaces.} To circumvent the need to compute expensive gradients, AS variants have been proposed. A simple but powerful idea for problems without a severe curse of dimensionality is to use the cheap derivatives of the 
posterior mean of a high-dimensional GP that approximates $f(\mathbf{x})$. This approach, known as OK-AS \cite{Palar2018}, can be seen as a special case of the gradient Kernel Dimension Reduction method. Indeed, OK-AS is identical to a version of gKDR where a linear 
kernel $k(f(\mathbf{x}), f(\mathbf{x'}))=f(\mathbf{x})^Tf(\mathbf{x'})$ is specified for the output space in (\ref{eq:gkdr_fukumizu}). The surrogate-assisted active subspace (SAAS) method \cite{Wycoff2021}  also expands on OK-AS. When estimating the AS, SAAS additionally 
incorporates the posterior uncertainty of a high‑dimensional GP surrogate that approximates $f(\mathbf{x})$. Although SAAS has proven its utility for dimension reduction and for improving GP modelling \cite{Wycoff2022, Binois2024}, one caveat is that the GP posterior 
uncertainty can be difficult to calibrate in high-dimensions \cite{seshadri2019dimension}. In terms of existing applications, both OK-AS and SAAS were implemented using anisotropic kernel functions, when the large number of hyperparameters was estimated using marginal likelihood
maximization. In contrast, gKDR has been applied using isotropic kernels with a small number of hyperparameters trained via cross-validation.\\

\subsection{New method: Sequential Ordinary Kriging Active Subspace (seqOK-AS)}\label{sec:seqokas_introduced}
The original OK-AS method \cite{Palar2018}, described in algorithm \ref{alg:SAAS}, consists of two steps. The first step is to use a GP surrogate for $f(\mathbf{x})$ to estimate the matrix $\mathbf{H}$ (\ref{eq:as_with_grad}). 
The second step is to use the resulting active subspace estimate to potentially improve the GP surrogate. Our idea is to repeat this two-step process iteratively, so that both the active subspace estimation and the GP surrogate approximation are improved.
\begin{algorithm}[H]
	\caption{GP surrogate improvement via gradient-free active subspace (OK-AS)}
	\label{alg:SAAS}
	\begin{algorithmic}[1]
        \State Construct a GP $F_n(\mathbf{x})$ using ordinary kriging with a training dataset $\{\mathbf{x}_i, f(\mathbf{x}_i)\}_{i=1}^{n}$
        \State Construct the estimated gradient covariance matrix using the kriging posterior mean $m_n$:
        $$\mathbf{H}_n=\int  \nabla m_n(\mathbf{x})\nabla m_n(\mathbf{x})^T q(\mathbf{x})d\mathbf{x}$$
        \State Extract the $r\leq d$ dominant eigenvectors $\mathbf{W}_n$ of $\mathbf{H}_n$ and project the data: $\{\mathbf{W}_n^T\mathbf{x}_i\}_{i=1}^n$
        \State Construct a new GP $F_{n,r}(\mathbf{W}_n^T\mathbf{x})$ using the estimate $\mathbf{W}_n$ and the data $\{\mathbf{W}_n^T\mathbf{x}_i, f(\mathbf{x}_i)\}_{i=1}^{n}$
        \State Compare the old GP surrogate $F_n(\mathbf{x})$ versus the new GP surrogate $F_{n,r}(\mathbf{W}_n^T\mathbf{x})$ on a validation set.
	\end{algorithmic}
\end{algorithm}
\par Indeed, our method seqOK-AS (Algorithm \ref{alg:seqSAAS}) is built upon the following two key observations from \cite{Palar2018}. First, for problems without a severe curse of dimensionality, the active subspace from (\ref{eq:as_with_grad}) can be estimated reasonably 
well using gradients of a high-dimensional GP posterior mean that approximates $f(\mathbf{x})$ (Algorithm \ref{alg:SAAS}, Step 2). Second, the GP approximation performance can be improved significantly by projecting the training inputs using the estimated active subspace 
onto a new coordinate system that is better aligned with the true active subspace (Algorithm \ref{alg:SAAS}, Step 5). 
\par Our first contribution is to observe that the improved GP surrogate resulting from Algorithm \ref{alg:SAAS} can be used to re-estimate the active subspace. Indeed, using gradients of the improved GP surrogate together with a simple application of the chain rule, 
one can go from the projected space back to the original coordinate system, and hence re-estimate the active subspace (Algorithm \ref{alg:seqSAAS}, Step 6:3). The improved GP performance should also translate to an improved active subspace estimate. Our second contribution
is to notice that one can use the re-estimated active subspace (Algorithm \ref{alg:seqSAAS}, Step 7:4) to project the training inputs onto a new coordinate system (Algorithm \ref{alg:seqSAAS}, Step 4:1). Thus, the GP performance might be further improved, given that the 
re-estimated active subspace should be better aligned with the true active subspace. This two-step process (i.e., re-estimate the active subspace and re-train the GP in the new coordinate system) comes with no extra simulation cost, and it is repeated iteratively until
performance improvements are no longer noticed. The performance improvements brought by our iterative process are confirmed by the numerical experiments from Section \ref{sec:surrogate_validation_numerical_tests}. On the downside, if the very first GP models are of poor 
quality, this can introduce a bias in the subsequent estimations. A design of experiments of sufficient size is necessary to minimize this bias. To assess the quality of the very first GP models, one can test the GP posterior mean via e.g., cross-validation. 
\par In principle, our approach seqOK-AS allows for a different truncation $r_k\leq d$ at each iteration (Algorithm \ref{alg:seqSAAS}, Step 7:4). This is motivated by the truncation strategy $r\leq d$ proposed for OK-AS (Algorithm \ref{alg:SAAS}, Step 3), which brought
performance improvements in some of the numerical experiments of \cite{Palar2018}. Nonetheless, \cite{Palar2018} shows that in most of the numerical tests, the performance improvements brought by OK-AS versus a high-dimensional GP surrogate in the original coordinate
space (Algorithm \ref{alg:SAAS}, Step 5) come from \textit{pure rotation} (i.e., $r=d$ in Algorithm \ref{alg:SAAS}, Step 3). In this vein, we did not observe any significant improvements due to truncations in our numerical tests for seqOK-AS. Thus, we use a sequence
of pure rotations for seqOK-AS (i.e., $r_k=d$ for all $k$ in  Algorithm \ref{alg:seqSAAS}, Step 7:4) in all our numerical experiments of Sec. \ref{sec:numerical_results}-\ref{sec:method_comparisonzz}. We conjecture that a truncation strategy could work better for problems 
with larger training budgets than considered in this work, since the active subspace estimation accuracy could be further improved. Indeed, low-dimensional GP surrogates trained on $\{\mathbf{\hat{W}}_{AS}^T\mathbf{x}_i, f(\mathbf{x}_i)\}_{i=1}^{n}$ for a gradient-based 
active subspace estimate $\mathbf{\hat{W}}_{AS}\in\mathbb{R}^{d\times r}$ with $r\ll d$ have shown great predictive performance \cite{Constantine2014, Palar2018, parente2020active}. Note that a large training budget may increase the performance of OK-AS to a level where 
seqOK-AS can offer little to no further improvement in terms of active subspace estimation or predictive performance. 
\par \ValentinCorrections{It is also worth noting that the version of Algorithm \ref{alg:seqSAAS} implemented using rotations alone can still be effective for dimensionality reduction.
Specifically, a gradient-free active subspace may be extracted by selecting the dominant $r<d$ eigenvectors $\mathbf{W}_K[:, 1:r]$ of the final estimate of the gradient covariance matrix, $\mathbf{H}_K$. 
This approach is employed in Sec. \ref{sec:numerical_results}, where seqOK-AS is evaluated against other gradient-free methods for active subspace estimation.}
\begin{algorithm}[H]
	\caption{Sequential Ordinary Kriging Active Subspace (\textbf{seqOK-AS})}
	\label{alg:seqSAAS}
	\begin{algorithmic}[1]
		\State Inputs: Training dataset $\{\mathbf{x}_i, f(\mathbf{x}_i)\}_{i=1}^{n}$, number of iterations $K$
		\State Initialization: Initial rotation matrix $\mathbf{W}_0=\mathbf{I}_d$, $r_0=d$
		\For{$k\in\{0,\cdots, K-1\}$}
			\State 1. Construct a GP $F_{n,k}$ with the data $\{\mathbf{W}_k^T\mathbf{x}_i, f(\mathbf{x}_i)\}_{i=1}^{n}$ using an automatic relevance determination (ARD) kernel (i.e., one lengthscale hyperparameter per dimension, $r_k$ lengthscales in total)
            \State 2. Train the hyperparameters of the GP $F_{n,k}$ using marginal likelihood maximization
			\State 3. Obtain the matrix $\mathbf{H}_{k+1}$ via the posterior mean $m_{n,k}$:
                 $$\mathbf{H}_{k+1}=\int  \nabla \tilde{m}_{n,k}(\mathbf{x})\nabla \tilde{m}_{n,k}(\mathbf{x})^T q(\mathbf{x})d\mathbf{x},$$
                 where $\tilde{m}_{n,k}(\mathbf{x}):=m_{n,k}(\mathbf{W}_k^T\mathbf{x})$ approximates $f(\mathbf{x})$ and $\nabla \tilde{m}_{n,k}(\mathbf{x})=\mathbf{W}_k\nabla m_{n,k}(\mathbf{W}_k^T\mathbf{x})$ using the chain rule
			\State 4. Let $\mathbf{W}_{k+1}\in\mathbb{R}^{d\times r_{k+1}}$ be the $r_{k+1}\leq d$ dominant eigenvectors of $\mathbf{H}_{k+1}$.
		\EndFor
	\end{algorithmic}
\end{algorithm}
\par \ValentinCorrections{The stability of Algorithm \ref{alg:seqSAAS} with respect to the number of iterations $K$ and the training size $n$ is demonstrated numerically in Sec. \ref{sec:numerical_results} (Fig. \ref{fig:rmse_fsa_quadratic_tests}-\ref{fig:rmse_fsa_physics_test}) and Appendix \ref{app:increasing_training_size_regression_benchmark} (Fig. \ref{fig:rmse_fsa_quadratic_comparison_increasing_training_size}-\ref{fig:rmse_fsa_real_world_comparison_increasing_training_size}), respectively, where a set of benchmark problems is considered. Providing a theoretical analysis of the proposed algorithm is a challenging task. Without additional assumptions on the target function $f$, one cannot expect meaningful convergence guarantees. It is therefore natural to relate our setting to existing frameworks in which structural assumptions on $f$ make the analysis tractable. A first line of research concerns the computation of active subspaces from approximate models. In uncertainty quantification and scientific computing, several papers have been devoted to understanding the robustness of active subspaces with respect to model approximations, see \cite{Constantine2015a,lam2020multifidelity,nobile2024function,geraci2019recent} (essentially, a perturbation analysis on the mapping $f\mapsto \mathbf{H}\mapsto \mathbf{W}_r$). A second line of research has recently attracted significant attention in machine learning and high-dimensional statistics. Here, one assumes that the target function admits a low-dimensional representation of the form
$f(\mathbf{x})=h\left((\mathbf{W}^\star)^\top \mathbf{x}\right)$, where $\mathbf{W}^\star\in\mathbb{R}^{d\times r}$ is an unknown projection matrix and $h:\mathbb{R}^r\to\mathbb{R}$ is an unknown link function. When $r=1$, this is the classical single-index model, introduced in the semiparametric statistics \cite{hardle1993optimal,ichimura1993semiparametric}. Closely related ideas appear in sufficient dimension reduction \cite{Li91SIR}. However, a rigorous theoretical analysis of the proposed iterative algorithm is beyond the scope of the present paper.}
\section{Application to high-dimensional rare event probability estimation}\label{sec:ce_seqsaas}
\noindent\textbf{The importance sampling framework.} Recall from Section \ref{sec:intro} that our reliability objective concerns the estimation of $P_f=\p[f(\mathbf{X})\leq 0]=\int \mathbb{I}[f(\mathbf{x})\leq 0] q(\mathbf{x})d\mathbf{x}$ for a random vector $\mathbf{X}$. The Importance Sampling (IS) estimator of this probability is
$$\hat{P}_f^{IS}=\frac{1}{N_{IS}}\sum_{i=1}^{N_{IS}}w(\mathbf{x}_i)\mathbb{I}[f(\mathbf{x}_i)\leq 0],$$
where $w(\mathbf{x}_i):=q(\mathbf{x}_i)/q_{IS}(\mathbf{x}_i)$ and the $\mathbf{x}_i$ are sampled from a biasing density $q_{IS}(\mathbf{x})$. The optimal IS biasing density which minimizes the variance of the estimator $\hat{P}_f^{IS}$ is
\begin{equation}\label{eq:optimal_is_density}
q^*_{IS}(\mathbf{x})\propto \mathbb{I}[f(\mathbf{x})\leq 0] q(\mathbf{x}) ,
\end{equation}
see \emph{e.g.,} \cite{rubinstein2016simulation}. In fact, this optimal IS estimator results in constant weights with zero variance. However, sampling from $q^*_{IS}$ is a difficult task, particularly because the support of $q^*_{IS}$, which is the failure domain $\mathcal{L}_f$ itself 
(\ref{eq:excursion_set}), is a priori unknown. \ValentinCorrections{Unfortunately, even if samples from $q^*_{IS}$ had been available, e.g., via MCMC methods \cite{au_estimation_2001}, the normalising constant in (\ref{eq:optimal_is_density}), needed for the IS weights, would still be unknown.} Our goal is to build a tractable low-dimensional approximation to $q^*_{IS}$ of the form 
$$
 q_{IS}(\mathbf{x}) = h(\mathbf{W_r}^\top \mathbf{x}) q(\mathbf{x}) ,
$$
where $\mathbf{W_r}\in\mathbb{R}^{d\times r}$ is a dimension reduction matrix with orthogonal columns, and $ h:\mathbb{R}^r \rightarrow \mathbb{R}$ is a low-dimensional positive function to be determined. 
In this section, we propose to identify $\mathbf{W_r}$ with the help of our new gradient-free method seqOK-AS (Algorithm \ref{alg:seqSAAS}), and to seek $ q_{IS}$ as a Gaussian density.
\par Note that sampling from $q^*_{IS}(\mathbf{x})$ is equivalent to solving a Bayesian inverse problem, if we consider $q(\mathbf{x})$ as the prior density and $\mathbb{I}[f(\mathbf{x})\leq 0]$ as the likelihood function. 
Consequently, methods that have shown good performance in various high-dimensional Bayesian inverse problems have been adapted to importance sampling for rare event probability estimation. Table \ref{Tab:table_bips_to_is} presents some of these methods. 
Conversely, reliability methods such as Subset Simulation \cite{au_estimation_2001} and Cross-Entropy based importance sampling with failure-informed dimension reduction \cite{Uribe2021cross} have been adapted to solve more general Bayesian inverse problems 
in the works of \cite{Straub2015bayesian} and \cite{Ehre2023certified}, respectively. 
\begin{table}[ht]
\caption{Bayesian inverse problem methods adapted to high-dimensional rare event reliability}
\centering
\vspace{0.5cm}
\begin{tabular}{||c c||} 
 \hline
 Original Method & High-dimensional reliability version \\ [0.5ex] 
 \hline\hline
 Sequential Monte Carlo \cite{Del2006sequential} & \cite{Papaioannou2016sequential} \\
 \hline
 Ensemble Kalman Inversion \cite{Iglesias2018bayesian} & \cite{Wagner2022ensemble}  \\
 \hline
 Consensus-Based Sampling \cite{Carrillo2022consensus} & \cite{Althaus2024consensus} \\
 \hline
\end{tabular}
\label{Tab:table_bips_to_is}
\end{table}

\noindent\textbf{Strategy via smooth approximation and dimension reduction.} Gradient-based methods have been proposed to identify the matrix $\mathbf{W_r}$ that captures the change from a reference density $q$ to a target density in the form 
$p(\mathbf{x}) \propto L(\mathbf{x})q(\mathbf{x})$ for some likelihood function $L:\mathbb{R}^{d}\rightarrow\mathbb{R}$. In particular in \cite{zahm2022certified,li2024principal,li2025sharp}, it is shown that defining $\mathbf{W_r}$ as the matrix containing
the dominant eigenvectors of the Fisher information matrix
\begin{equation}\label{eq:H_CDR}
\mathbf{H} = \int \nabla\ln L(\mathbf{x}) \nabla\ln L(\mathbf{x})^\top p(\mathbf{x})d\mathbf{x},
\end{equation}
permits to control various divergences between the density of interest $p$ and a low-dimensional approximation of it in the form of $\widetilde p(\mathbf{x}) = h(\mathbf{W_r}^\top \mathbf{x}) q(\mathbf{x})$. \ValentinCorrections{Specifically, \cite{cui2022unified,li2024principal} proved that, for some optimal choice for the function $h:\R^r\rightarrow\R_+$, the Hellinger distance $D_\mathrm{Hell}( \widetilde p, p ) = (\frac{1}{2}\int ( \sqrt{\widetilde p} - \sqrt{p} )^2 d\mathbf{x} )^{1/2}$ is bounded as follows:
\begin{equation}\label{eq:CDR_hellinger}
 D_\mathrm{Hell}( \widetilde p, p )^2 \leq \frac{C_\mathrm{PI}(q)}{4} \left( \tr(\mathbf{H}) - \tr( \mathbf{W_r}^\top\mathbf{H}\mathbf{W_r} ) \right),
\end{equation}
where $C_\mathrm{PI}(q)$ is the subspace Poincar\'e constant of $q$.} However, this framework does not directly apply to 
the optimal IS biasing $p=q^*_{IS}$ (\ref{eq:optimal_is_density}) because the function $L(\bf x)=\mathbb{I}[f(\mathbf{x})\leq 0]$ is not regular enough for the method to apply.
\par Following \cite{Uribe2021cross}, we consider one additional smoothing step, which consists of approximating the target $q^*_{IS}$ with
\begin{equation}\label{eq:smoothing_is}
 q^\varepsilon_{IS} \propto h^\varepsilon( - f(\mathbf{x}) ) q(\mathbf{x}) ,
\end{equation}
where $h^\varepsilon$ is a smooth sigmoid approximation to the Heaviside function $ t\mapsto \mathbb{I}[ t\geq0 ] $. In other words, as $\varepsilon\to 0$, $h^\varepsilon(-f(\mathbf{x}))$ approximates $\mathbb{I}[f(\mathbf{x})\leq 0]$. Smoothing functions have also been used in the context of Active Subspace, see \cite{verdiere2025mollified}.
For instance, we can use the following approximation defined as the square of the cumulative distribution function (CDF) of a normal distribution:
\begin{equation}\label{eq:h_CDF}
h^\varepsilon(t) = \left( \frac{1}{ 2\varepsilon \sqrt{\pi} } \int_{-\infty}^t  \exp\left(-\frac{z^2}{4\varepsilon^2}\right) d z \right)^2
\quad\longrightarrow \quad
\begin{cases}
 0 &\text{as } t\rightarrow -\infty \\
 1 &\text{as } t\rightarrow +\infty
\end{cases} .
\end{equation}
With this choice, the matrix $\mathbf{H^\varepsilon}$ as defined in \eqref{eq:H_CDR} with $L^\varepsilon(\mathbf{x}) = h^\varepsilon( - f(\mathbf{x}) )$ becomes
\begin{align} 
 \mathbf{H^\varepsilon}
 &= \int \nabla\ln L^\varepsilon(\mathbf{x}) \nabla\ln L^\varepsilon(\mathbf{x})^\top \frac{ L^\varepsilon(\mathbf{x}) q(\mathbf{x})}{\int L^\varepsilon(\mathbf{x'})q(\mathbf{x'}) d\mathbf{x'}}d\mathbf{x}\label{eq:active_subspace_reliability_formula} \\
 &\overset{\eqref{eq:h_CDF}}{=}
 \frac{1}{\varepsilon^2 \pi P_f^\varepsilon  }\int \nabla f(\mathbf{x}) \nabla f(\mathbf{x})^\top   \exp\left(-\frac{f(\mathbf{x})^2}{2\varepsilon^2}\right) q(\mathbf{x})   d\mathbf{x},
\end{align}
where $P_f^\varepsilon = \int L^\varepsilon(\mathbf{x}) q(\mathbf{x}) d\mathbf{x}$ is the normalizing constant of $q^\varepsilon_{IS}$.
Remarkably, this matrix $\mathbf{H^\varepsilon}$ is similar to the Active Subspace matrix as defined in \eqref{eq:as_with_grad}, the main difference being that the expectation is taken over $q_\varepsilon(\mathbf{x})\propto \exp(-\frac{f(\mathbf{x})^2}{2\varepsilon^2}) q(\mathbf{x}) $ instead of over $q(\mathbf{x})$.
As the smoothing parameter $\varepsilon\rightarrow 0$ goes to zero, the mass of $q_\varepsilon$ concentrates around the boundary of the failure region.
In other words, when using the AS method for rare event probability estimation, the transition region between $f(\mathbf{x})\geq0$ and  $f(\mathbf{x})\leq0$ is the relevant region for sampling the gradient. \ValentinCorrections{This remark is the key motivation for using the \emph{goal-oriented} approach, which consists in approximating $  L^\varepsilon(\mathbf{x})$ instead of $f(\mathbf{x})$ directly.}\\
\noindent\textbf{Towards an algorithm.} \ValentinCorrections{Since different smoothing approximations will be discussed in Sec. \ref{sec:method_comparisonzz}, let us write the Fisher information matrix $\mathbf{H}^{\varepsilon}$ for a generic smoothing function $h^\varepsilon$ with $L^\varepsilon(\mathbf{x}) = h^\varepsilon( - f(\mathbf{x}) )$ as:
\begin{equation}\label{eq:Fisher_info_generic}
    \mathbf{H^\varepsilon}
 = \int \nabla\ln L^\varepsilon(\mathbf{x}) \nabla\ln L^\varepsilon(\mathbf{x})^\top \frac{ L^\varepsilon(\mathbf{x}) q(\mathbf{x})}{\int L^\varepsilon(\mathbf{x'})q(\mathbf{x'}) d\mathbf{x'}}d\mathbf{x}.
\end{equation}}
Inspired by \cite{Uribe2021cross}, we propose using a sequential approach with a sequence of importance densities parametrized by decreasing temperatures $\infty=\varepsilon_0\geq\varepsilon_1\geq \dots\geq\varepsilon_m \approx 0$. In this regard, a collection of particles starts from $q$ ($\varepsilon_0=\infty$) and then sequentially approaches an approximation of $p=q^*_{IS}$ (\ref{eq:optimal_is_density}) ($\varepsilon_m \approx 0$).  For each step $j\geq 1$, we use importance sampling with the density from the previous step $\tilde{\pi}^{\varepsilon_{j-1}, r_{j-1}}$ to estimate the matrix $\mathbf{H}^{\varepsilon_j}$ (\ref{eq:Fisher_info_generic}). Then, we find the best Gaussian approximation $\tilde{\pi}^{\varepsilon_j, r_j}$ of the new projected density $\pi^{\varepsilon_j,r_j}(\mathbf{x}) \propto \tilde{L}^{\varepsilon_j}( \mathbf{\hat{W}}_{r_j}^\top \mathbf{x} ) q(\mathbf{x})$ via a low-dimensional cross-entropy minimization, denoted \ValentinCorrections{iCEred} \cite{Uribe2021cross}. 
\ValentinCorrections{The temperature $\varepsilon_j$ is chosen such that the distance between consecutive densities $\tilde{\pi}^{\varepsilon_{j-1}, r_{j-1}}(\mathbf{x})$ and $\pi^{\varepsilon_j}(\mathbf{x})\propto L^{\varepsilon_j}(\bs{x})q(\bs{x})$ is controlled \cite{Papaioannou2016sequential}; thus, the IS weights do not degenerate as $j$ increases.}
In this context, our contribution is to use our strategy seqOK-AS to avoid the calculation of gradients $\nabla L^{\varepsilon_j}(\mathbf{x})$ when estimating $\mathbf{H}^{\varepsilon_j}$.
Gradients were previously used by \cite{Uribe2021cross} in all the numerical tests.
\par \ValentinCorrections{For a fixed computational cost and for sufficiently large $d$, the CE method with dimension reduction proposed by \cite{Uribe2021cross} can significantly outperform an analogue approach of CE minimization without dimension reduction in terms of failure probability estimation. In other words, dimension reduction can play a key role in keeping the number of particles low. Further, \cite{tong2023large} demonstrates that for a high-dimensional quadratic model and a fixed computational cost, the iCEred method outperforms the classic approach of Importance Sampling using FORM initialization \cite{schueller_critical_1987}.}\\
\noindent\textbf{Proposed procedure for estimating the probability of rare events (\textit{iCEred+seqOK-AS}).}
Within the loop involving the different importance sampling densities, we propose to integrate our seqOK-AS algorithm. Thus, for $j=1,2,\dots,\max_{\textrm{iter}}+1$:
\begin{itemize}
     \item[$\circ$] For $i=[1, N]$, we sample points $\mathbf{x}_i^{(j-1)}\sim\tilde{\pi}^{\varepsilon_{j-1}, r_{j-1}}$ and evaluate $f(\mathbf{x}_i^{(j-1)})$ (recall that $\tilde{\pi}^{\varepsilon_{0}, r_{0}}(\mathbf{x})=q(\mathbf{x})$),
     \item[$\circ$] If $j=\max_{\textrm{iter}}+1$, then the iCEred+seqOK-AS algorithm terminates at step $m:=j-1=\max_{\textrm{iter}}$; otherwise, if $j\leq \max_{\textrm{iter}}$, then proceed to the next instruction below,
     \item[$\circ$] \ValentinCorrections{We check the early stopping criterion, i.e., if the distance between $q^*_{IS}(\mathbf{x})\propto \mathbb{I}[f(\mathbf{x})\leq 0] q(\mathbf{x})$ and $\pi^{\varepsilon_{j-1}}(\mathbf{x})\propto h^{\varepsilon_{j-1}}(-f(\bs{x}))q(\bs{x})$ is small enough based on the following sample coefficient of variation test (for a square-integrable random variable $U$, the coefficient of variation is defined as $\text{CoV}[U]:=|\E[U]|/\sqrt{\textrm{Var}[U]}$):
     \begin{equation}\label{eq:stopping_criterion_is}
     \widehat{\text{CoV}}\Bigg(\Bigg\{\frac{\mathbb{I}[f(\mathbf{x}_i^{(j-1)})\leq 0]q(\bs{x}_i^{(j-1)})}{h^{\varepsilon_{j-1}}(-f(\bs{x}_i^{(j-1)}))q(\bs{x}_i^{(j-1)})}\Bigg\}_{i=1}^N\Bigg)=\widehat{\text{CoV}}\Bigg(\Bigg\{\frac{\mathbb{I}[f(\mathbf{x}_i^{(j-1)})\leq 0]}{h^{\varepsilon_{j-1}}(-f(\bs{x}_i^{(j-1)}))}\Bigg\}_{i=1}^N\Bigg)\leq\delta
     \end{equation}
     then the iCEred+seqOK-AS algorithm terminates at step $m:=j-1$; otherwise, if:
     $$\widehat{\text{CoV}}\Bigg(\Bigg\{\frac{\mathbb{I}[f(\mathbf{x}_i^{(j-1)})\leq 0]}{h^{\varepsilon_{j-1}}(-f(\bs{x}_i^{(j-1)}))}\Bigg\}_{i=1}^N\Bigg)>\delta$$
     then proceed to the next instruction below,}
     \item[$\circ$] \ValentinCorrections{The temperature $\varepsilon_j$ is selected such that the following sample coefficient of variation is kept close to a constant value $\delta$:
     $$\varepsilon_j=\argminE_{\varepsilon\in(0, \varepsilon_{j-1})}\Bigg(\widehat{\text{CoV}}\Bigg(\Bigg\{\frac{h^{\varepsilon}(-f(\bs{x}_i^{(j-1)}))q(\bs{x}_i^{(j-1)})}{\tilde{\pi}^{\varepsilon_{j-1}, r_{j-1}}(\bs{x}_i^{(j-1)})}\Bigg\}_{i=1}^N\Bigg)-\delta\Bigg)^2;$$
     this ensures that the distance between $\pi^{\varepsilon_j}(\mathbf{x})\propto h^{\varepsilon_j}(-f(\bs{x}))q(\bs{x})$ and $\tilde{\pi}^{\varepsilon_{j-1}, r_{j-1}}(\mathbf{x})$ is not too large,}
     \item[$\circ$] We form a training set with all points evaluated so far, i.e., $(\mathbf{x}_i^{(k)}, L^{\varepsilon_j}(\mathbf{x}_i^{(k)}))$ for $i=1,...,N$ and $k=0,...,j-1$; this is immediate, since $L^{\varepsilon_j}(\mathbf{x})= h^{\varepsilon_j}( - f(\mathbf{x}) )$ and  $(\mathbf{x}_i^{(k)}, f(\mathbf{x}_i^{(k)}))$ for $i=1,...,N$ and $k=0,...,j-1$ are available, 
     \item[$\circ$] We train a GP metamodel for $L^{\varepsilon_j}$ and estimate its gradient $\nabla L^{\varepsilon_j}(\mathbf{x})$ using seqOK-AS (Algorithm \ref{alg:seqSAAS}); \ValentinCorrections{the only purpose served by the GP surrogate is to estimate the gradient,}
     \item[$\circ$] The gradient of the metamodel's posterior mean, $\nabla \tilde{m}_{j,K}(\mathbf{x})
     $, allows for a gradient-free estimation of $\mathbf{H}^{\varepsilon_j}$  (\ref{eq:Fisher_info_generic}) via self-normalized IS:
     $$\hat{\mathbf{H}}^{\varepsilon_j}=\frac{1}{\bar{w}^{(j)}}\sum_{i=1}^N\tilde{w}_i^{(j)}[\nabla\log\tilde{m}_{j,K}(\mathbf{x}_i^{(j-1)})][\nabla\log\tilde{m}_{j,K}(\mathbf{x}_i^{(j-1)})]^\top,$$
     where $\nabla\log\tilde{m}_{j,K}(\mathbf{x})$ is obtained from $\nabla\tilde{m}_{j,K}(\mathbf{x})$ using the chain rule and the weights $\tilde{w}_i^{(j)}$ are given by: 
    $$
    \tilde{w}_i^{(j)}=\frac{L^{\varepsilon_j}(\bs{x}_i^{(j-1)})q(\bs{x}_i^{(j-1)})}{\tilde{\pi}^{\varepsilon_{j-1}, r_{j-1}}(\bs{x}_i^{(j-1)})}
    $$ 
     and $\bar{w}^{(j)}=\sum_{i=1}^N \tilde{w}_i^{(j)}$; \ValentinCorrections{the weights are computed using the available model evaluations $\{(\mathbf{x}_i^{(j-1)}, L^{\varepsilon_j}(\mathbf{x}_i^{(j-1)}))\}_{i=1}^N$,}
     \item[$\circ$] Let $\mathbf{\hat{W}}_{r_j}\in\R^{d\times r_j}$ be the matrix whose columns are the dominant $r_j\leq d$ eigenvectors of $\hat{\mathbf{H}}^{\varepsilon_j}$, \ValentinCorrections{i.e., $r_j=\min\{r':\frac{1}{2}\sum_{l=r'+1}^d\lambda_l\leq\varepsilon_{AS}\}$, where $\lambda_1\geq\cdots\geq\lambda_l\geq\cdots\lambda_d\geq0$ are the eigenvalues of $\hat{\mathbf{H}}^{\varepsilon_j}$,}
     \item[$\circ$] The best Gaussian approximation $\tilde{\pi}^{\varepsilon_{j}, r_{j}}(\mathbf{x})$ of the projected density $\pi^{\varepsilon_j,r_j}(\mathbf{x}) \propto \tilde{L}^{\varepsilon_j}( \mathbf{\hat{W}}_{r_j}^\top \mathbf{x} ) q(\mathbf{x})$ is sought using the efficient Cross-Entropy minimization method in the low-dimensional space $r_j\leq d$; \ValentinCorrections{given the choice of a standard Gaussian density for $q(x)$, obtained possibly after an isoprobabilistic transformation such as Rosenblatt or Nataf, a closed-form solution for $\tilde{\pi}^{\varepsilon_{j}, r_{j}}(\mathbf{x})$ is obtained cheaply using the projection matrix $\mathbf{\hat{W}}_{r_j}$ and model evaluations $\{(\mathbf{x}_i^{(j-1)}, L^{\varepsilon_j}(\mathbf{x}_i^{(j-1)}))\}_{i=1}^N$ (see \cite{Uribe2021cross} for full details),}
     \item[$\circ$] Proceed to the next iteration $j\leftarrow j+1$.
\end{itemize}
\par The final Gaussian approximation $\tilde{\pi}^{\varepsilon_{m}, r_{m}}(\mathbf{x})$ is used for estimating $P_f$ via Importance Sampling:
\begin{equation}\label{eq:final_is_step}
\widehat{P}_f^{IS}=\frac{1}{N_{IS}}\sum_{i=1}^{N_{IS}}\tilde{w}(\mathbf{x}_i)\mathbb{I}[f(\mathbf{x}_i)\leq 0],
\end{equation}
where $\tilde{w}(\mathbf{x}_i):=q(\mathbf{x}_i)/\tilde{\pi}^{\varepsilon_{m}, r_{m}}(\mathbf{x})$ and the $\mathbf{x}_i$ are sampled from the biasing density $\tilde{\pi}^{\varepsilon_{m}, r_{m}}(\mathbf{x})$. \ValentinCorrections{Two options for the choice of $N_{IS}$ are proposed by \cite{Uribe2021cross}, which are also followed in this work. First, the straightforward option $N_{IS}=N$ using the data already available: $\{(\mathbf{x}_i^{(m)}, f(\mathbf{x}_i^{(m)}))\}_{i=1}^N$. Second, an alternative option $N_{IS}>N$ where additional samples $(\mathbf{x}_{N+j}\sim\tilde{\pi}^{\varepsilon_{m}, r_{m}}(\mathbf{x}), f(\mathbf{x}_{N+j}))$ are iteratively added to $\{(\mathbf{x}_i^{(m)}, f(\mathbf{x}_i^{(m)}))\}_{i=1}^N$ until the stopping criterion based on the coefficient of variation for the probability of failure is reached:
\begin{equation}\label{eq:stop_criterion_refinement_IS}
\widehat{\text{CoV}}(\widehat{P}_f^{IS}):=\frac{\sqrt{\widehat{\text{Var}}[\widehat{P}_f^{IS}]}}{\widehat{P}_f^{IS}}\leq\overline{\delta},
\end{equation}
where $\widehat{\text{Var}}[\widehat{P}_f^{IS}]:=\frac{1}{N_{IS}-1}\Big[\frac{1}{N_{IS}}\sum_{i=1}^{N_{IS}}\tilde{w}(\mathbf{x}_i)^2\mathbb{I}[f(\mathbf{x}_i)\leq 0]-(\widehat{P}_f^{IS})^2\Big]$ and $\overline{\delta}$ is an additional tuning parameter.}
\par \ValentinCorrections{The iCEred+seqOK-AS procedure is numerically validated in Sec. \ref{sec:method_comparisonzz}, where most of the parameters follow the guidelines set for iCEred by \cite{Uribe2021cross}. 
Namely, a constant number of particles $N$ is used for each iCEred+seqOK-AS iteration. The threshold for the stopping criterion (\ref{eq:stopping_criterion_is}) of iCEred+seqOK-AS is fixed at $\delta=1.5$, as also suggested by \cite{Papaioannou2016sequential}. 
The tolerance for the active subspace truncation is fixed at $\varepsilon_{AS}=0.01$. The refinement option $N_{IS}>N$ for the estimator $\widehat{P}_f^{IS}$ is triggered only once in the experiments of Sec. \ref{sec:method_comparisonzz}, where a threshold $\overline{\delta}=0.2$ is used for the stopping criterion (\ref{eq:stop_criterion_refinement_IS}). This choice deviates from the more conservative option $\overline{\delta}=0.05$ suggested by \cite{Uribe2021cross}. 
It is also important to mention that in some of the numerical experiments of Sec. \ref{sec:method_comparisonzz}, the GP surrogates are given a warm start of $N_{init}>0$ training points $\{\mathbf{x}_i, L^{\varepsilon_j}(\mathbf{x}_i)\}_{i=1}^{N_{init}}$, where $\mathbf{x}_i\sim q$ and the corresponding model evaluations $\{\mathbf{x}_i, f(\mathbf{x}_i)\}_{i=1}^{N_{init}}$ are pre-computed before the start of the iCEred+seqOK-AS procedure. The parameter $N_{init}$ provides a trade-off that can help us reduce the number of particles $N$ subsequently needed for iCEred+seqOK-AS convergence.}
\par \ValentinCorrections{We finish this section with a different perspective. Recall that our proposed algorithm estimates $P_f$ as in (\ref{eq:final_is_step}), i.e., via IS based on a Gaussian approximation of $\pi^{\varepsilon_m,r_m}(\mathbf{x}) \propto \tilde{L}^{\varepsilon_m}( \mathbf{\hat{W}}_{r_m}^\top \mathbf{x} ) q(\mathbf{x})$. Alternatively, one could use a potentially different dimension reduction strategy $\mathbf{W}_r\in\R^{d\times r}$ coupled with a suitable low-dimensional approximation $\widetilde p^\varepsilon(\mathbf{x}) \propto h(\mathbf{W_r}^\top \mathbf{x}) q(\mathbf{x})$ of $p^\varepsilon(\mathbf{x}) \propto  h^\varepsilon( -f(\mathbf{x}) ) q(\mathbf{x})$ to estimate $P_f$ via the normalising constant
\begin{equation}\label{eq:alternative_Pf_estimation}
\widetilde P_f^{\varepsilon} = \int h( \mathbf{W}_{r}^\top \mathbf{x} ) q(\mathbf{x}) d\mathbf{x}.    
\end{equation}
By combining the smoothing of the Heaviside function with dimensionality reduction, we introduce two approximation steps. The following proposition demonstrates how these two sources of error can be optimally balanced from the perspective of computing rare event probability. The proof is given in Appendix \ref{app:proposition_smoothing_dimension_reduction_balance}.}
\begin{prpstn}\label{prop:alternative_Pf_estimation}
\ValentinCorrections{Let $f:\R^d\rightarrow\R$ be a smooth function and $q$ a probability density.
Consider the probability densities $p(\mathbf{x}) = \frac{1}{P_f} \mathbb{I}(f(\mathbf{x})\leq 0) q(\mathbf{x}) $ and $p^\varepsilon(\mathbf{x}) = \frac{1}{P_f^\varepsilon} h^\varepsilon( -f(\mathbf{x}) ) q(\mathbf{x}) $, where
$P_f = \int \mathbb{I}(f(\mathbf{x})\leq 0) q(\mathbf{x}) d\mathbf{x}$ and  $P_f^\varepsilon = \int h^\varepsilon( -f(\mathbf{x}) ) q(\mathbf{x}) d\mathbf{x}$ are normalising constants and where $h^\varepsilon$ is the smoothed Heaviside function as in \eqref{eq:h_CDF} for some $\varepsilon>0$. Then, for any matrix $\mathbf{W_r}\in\R^{d\times r}$ with orthogonal columns, there exists a function $h:\R^r\rightarrow\R$ such that the probability density $\widetilde p^\varepsilon(\mathbf{x}) = \frac{1}{\widetilde P_f^\varepsilon} h( \mathbf{W_r}^\top \mathbf{x} ) q(\mathbf{x})$ with normalising constant $\widetilde P_f^\varepsilon = \int h( \mathbf{W_r}^\top \mathbf{x} ) q(\mathbf{x}) d\mathbf{x}$ satisfies
\begin{equation}\label{eq:proposition_smoothing_dimension_reduction_balance}
 \left(\sqrt{P_f} - \sqrt{\widetilde P_f^\varepsilon}\right)^2
 \leq \underbrace{\frac{1}{2} \int \exp\left(-\frac{f(\mathbf{x})^2}{2\varepsilon^2}\right)  q(\mathbf{x}) d \mathbf{x}}_{A} +
  \underbrace{ P_f^\varepsilon C_\mathrm{PI}(q)  \left( \tr(\mathbf{H^\varepsilon}) - \tr( \mathbf{W_r}^\top\mathbf{H^\varepsilon}\mathbf{W_r} ) \right) }_{B}
\end{equation}
where $C_\mathrm{PI}(q)$ is the subspace Poincar\'e constant of $q$.}
\end{prpstn}
\ValentinCorrections{Notice that the term $A$ in (\ref{eq:proposition_smoothing_dimension_reduction_balance}) corresponds to the probability mass ``around'' the transition phase $f(\mathbf{x})\approx 0$. In particular, it becomes small when $\varepsilon\rightarrow 0$ (one can even show that $A=\mathcal{O}(\varepsilon)$ provided $\nabla f(\mathbf{x})\neq 0$ whenever $f(\mathbf{x})=0$). In turn, the term $B$ in (\ref{eq:proposition_smoothing_dimension_reduction_balance}) captures how well the dimension reduction can be performed in the same region $f(\mathbf{x})\approx 0$. This justifies that, up to an error of $\mathcal{O}(\varepsilon)$, the active subspace for a rare-event problem can be well captured by the diagnostic matrix $\mathbf{H^\varepsilon}$.}

\section{Numerical results for high-dimensional GP regression}\label{sec:numerical_results}
\subsection{Measures of quality and tested functions }\label{sec:surrogate_validation_numerical_tests}
We test our seqOK-AS approach (Algorithm \ref{alg:seqSAAS}) on various test problems using two metrics. First, the (normalised) root mean square error (RMSE) on a validation set of size $N$, i.e.,
$$
\frac{\left(\frac{1}{N}\sum_{i=1}^{N}(m_{n,K}(\mathbf{W}_K^T \mathbf{x}_i)-f(\mathbf{x}_i))^2\right)^{1/2}}{\max_{i=1:N} f(\mathbf{x}_i)-\min_{i=1:N} f(\mathbf{x}_i)}
$$
and second, the first subspace angle (FSA) between the seqOK-AS estimated and true active subspace: 
$$
\|\mathbf{W}_K[:,1:r]^T\mathbf{W}_{\text{true}}[:,1:r]^\perp\|_F,
$$
where $\mathbf{W}_{\text{true}}[:,1:r]$ are the $r$ dominant eigenvectors of $\mathbf{H}$ (\ref{eq:as_with_grad}), estimated using standard Monte Carlo with true gradients, $\mathbf{W}_{\text{true}}[:,1:r]^\perp\in\mathbb{R}^{d\times (d-r)}$ is the orthogonal complement 
of $\mathbf{W}_{\text{true}}[:,1:r]$, and $||.||_F$ is the Frobenius norm. For a matrix $\mathbf{A}$ with $\mathbf{A}^\star$ its conjugate transpose, the Frobenius norm is defined as $||\mathbf{A}||_F:=\sqrt{\tr(\mathbf{A}^\star \mathbf{A})}$. 
\par We use a set of models $f(\mathbf{x}):\mathbb{R}^d\to\mathbb{R}$ from the existing literature \cite{Gautier2022}. All the test problems exhibit an active subspace of dimension $r \in \{1, 2\}$. \ValentinCorrections{This observation is based on the eigenvalues $\lambda_1\geq\dots\geq \lambda_d\geq 0$ of $\mathbf{H}$ (\ref{eq:as_with_grad}), estimated using Monte Carlo with true gradients, according to the following selection criterion: $$r:=\min\{r'\in\{1,\dots,d-1\}:\lambda_{r'}\geq 10\lambda_{r'+1}\},$$ i.e., the $r^{th}$ eigenvalue of $\mathbf{H}$ is at least one order of magnitude larger than the next.
The size of each training set is $n=5d$, as in \cite{Gautier2022} (see Appendix \ref{app:increasing_training_size_regression_benchmark} for a discussion and results for increasing training size $n\in\{d, 2d, 3d, 4d, 5d, 6d, 7d\}$):}
\begin{itemize}[label=$\circ$]
     \item \textbf{Quadratic function:} $g(\mathbf{z})=\mathbf{z}^T\mathbf{A}
     \mathbf{z}+\mathbf{b}\mathbf{z}+c+\varepsilon$ with $\mathbf{z}=\mathbf{W}^T\mathbf{x}$, $\mathbf{x}\in\mathbb{R}^d$, $\mathbf{z}\in\mathbb{R}^r$. We study the cases $d\in\{25, 50, 100\}$ and $r\in\{1, 2\}$. Each element of $\mathbf{A}$, $\mathbf{b}$, and $c$ is independently sampled from the standard Gaussian $\mathcal{N}(0, 1)$, and $\varepsilon\sim\mathcal{N}(0, \sigma^2=25\cdot10^{-4})$ acounts for the variation in the inactive directions. 
     \item \textbf{Lift coefficient for a NACA0012 airfoil} \cite{Constantine2015a}: $d=18, r=1$.
     \item \textbf{Lift coefficient for an ONERA-M6 wing} \cite{Lukaczyk2014}: $d=50, r=1$.
     \item \textbf{HIV long-term model} \cite{Loudon2016}: $d=27, r=1$.
     \item \textbf{Elliptic PDE} \cite{Constantine2016accelerating}: $d=100, r=1$.
\end{itemize}
\par
\begin{figure}[htbp]
  \centering
  \begin{tabular}{ccc}
    \begin{subfigure}{0.31\textwidth}
      \includegraphics[width=\linewidth]{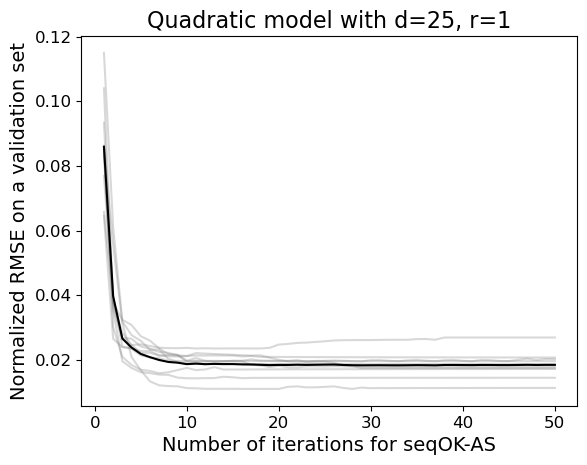}
    \end{subfigure} &
    \begin{subfigure}{0.31\textwidth}
      \includegraphics[width=\linewidth]{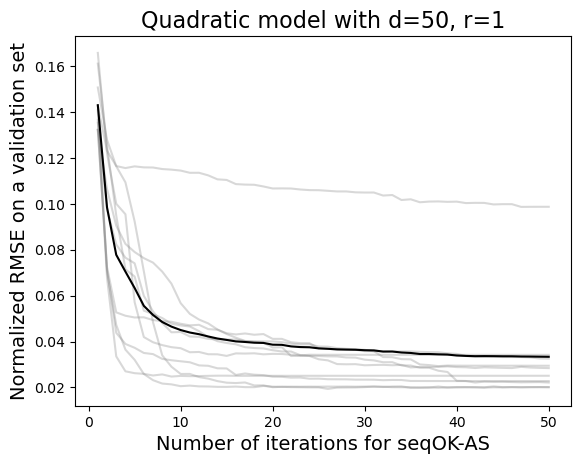}
    \end{subfigure} &
    \begin{subfigure}{0.31\textwidth}
      \includegraphics[width=\linewidth]{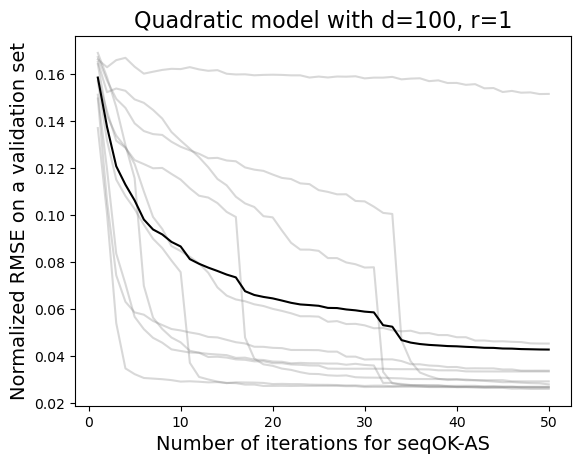}
    \end{subfigure} \\

    \begin{subfigure}{0.31\textwidth}
      \includegraphics[width=\linewidth]{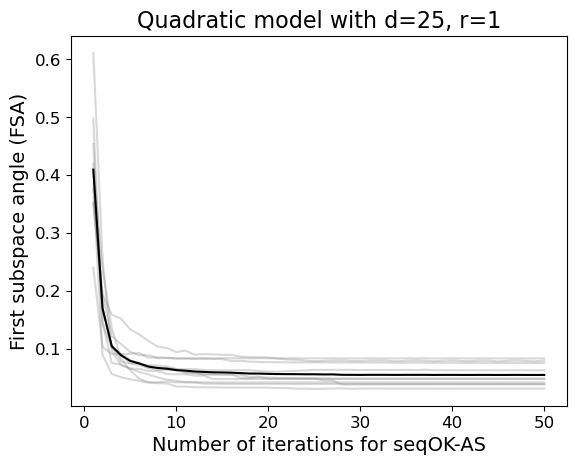}
    \end{subfigure} &
    \begin{subfigure}{0.31\textwidth}
      \includegraphics[width=\linewidth]{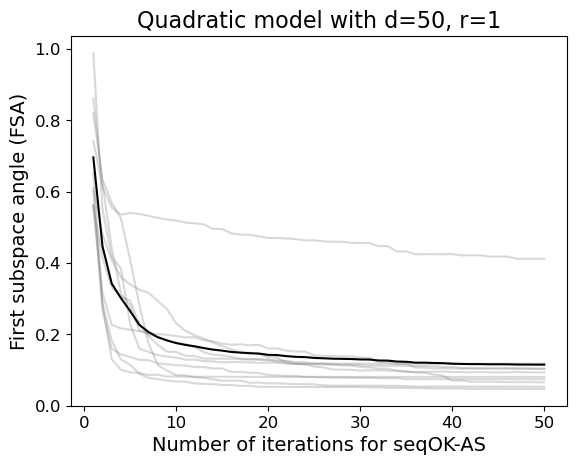}
    \end{subfigure} &
    \begin{subfigure}{0.31\textwidth}
      \includegraphics[width=\linewidth]{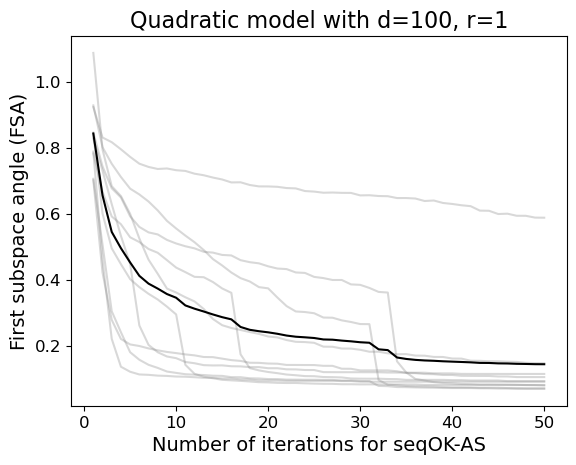}
    \end{subfigure} \\

    \begin{subfigure}{0.31\textwidth}
      \includegraphics[width=\linewidth]{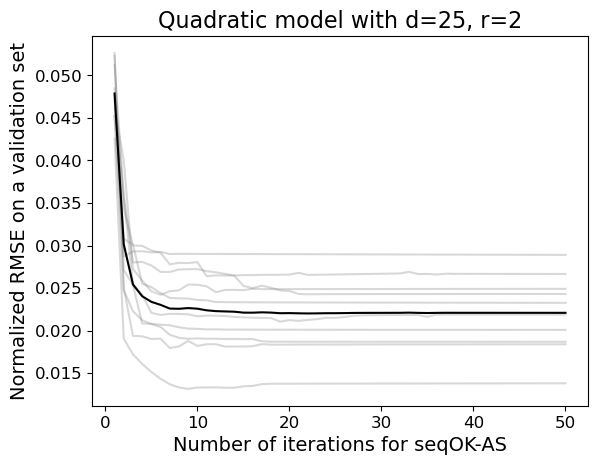}
    \end{subfigure} &
    \begin{subfigure}{0.31\textwidth}
      \includegraphics[width=\linewidth]{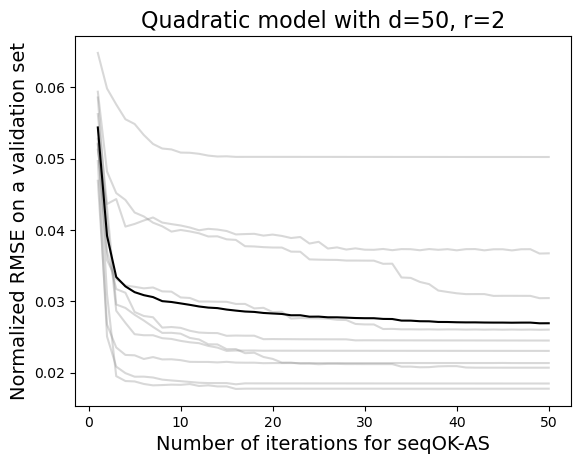}
    \end{subfigure} &
    \begin{subfigure}{0.31\textwidth}
      \includegraphics[width=\linewidth]{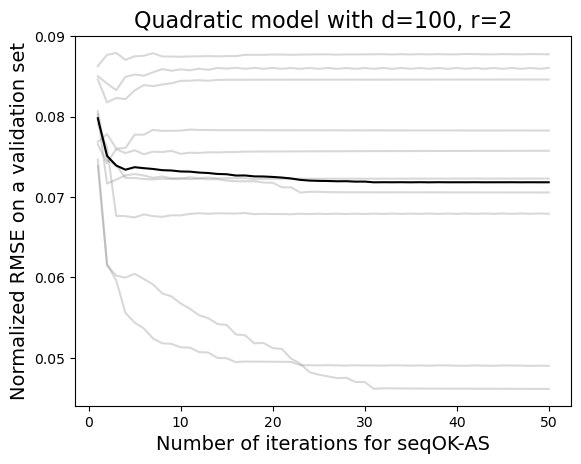}
    \end{subfigure} \\

    \begin{subfigure}{0.31\textwidth}
      \includegraphics[width=\linewidth]{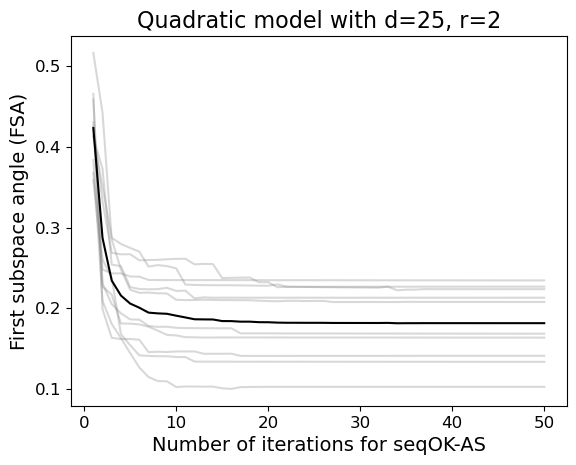}
    \end{subfigure} &
    \begin{subfigure}{0.31\textwidth}
      \includegraphics[width=\linewidth]{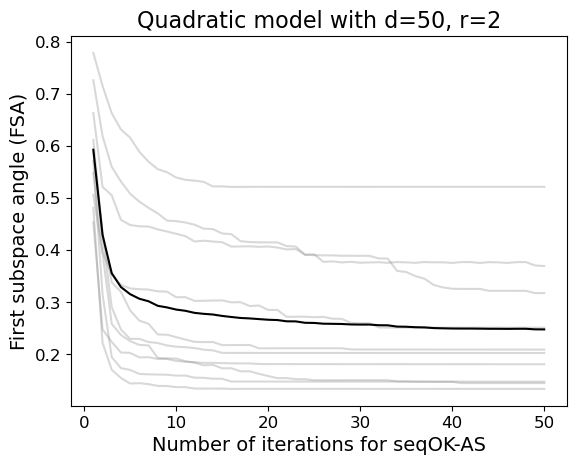}
    \end{subfigure} &
    \begin{subfigure}{0.31\textwidth}
      \includegraphics[width=\linewidth]{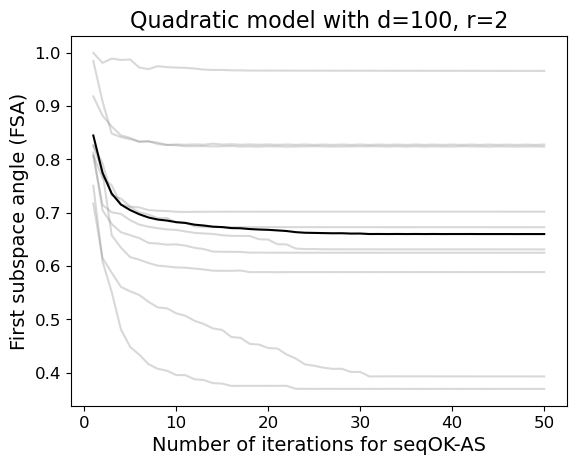}
    \end{subfigure}
  \end{tabular}
  \caption{Validation of the seqOK-AS method on the quadratic model (solid line: average over 10 trials)}
  \label{fig:rmse_fsa_quadratic_tests}
\end{figure}

\begin{figure}[htbp]
  \centering
  \begin{tabular}{cc}
    \begin{subfigure}{0.35\textwidth}
      \includegraphics[width=\linewidth]{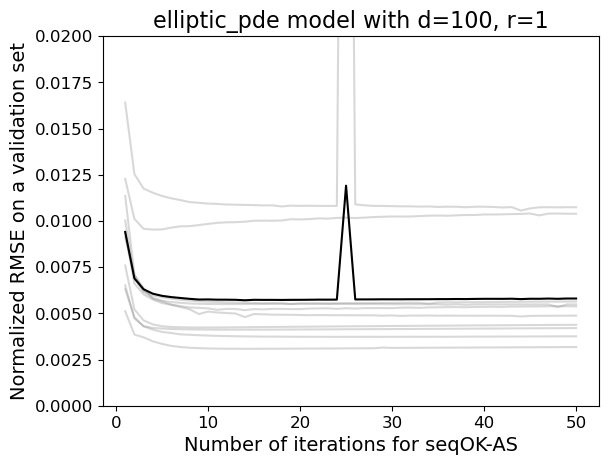}
    \end{subfigure} &
    \begin{subfigure}{0.35\textwidth}
      \includegraphics[width=\linewidth]{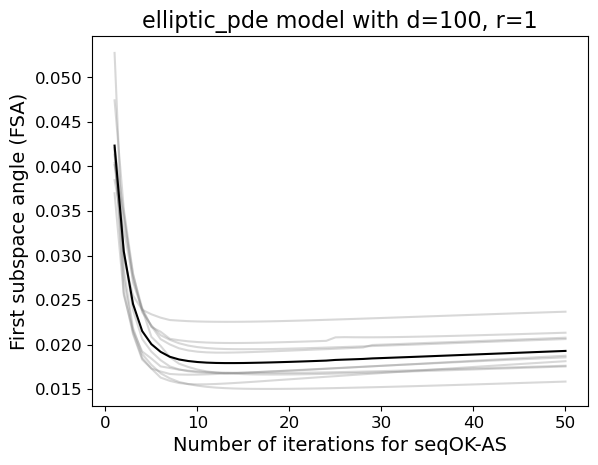}
    \end{subfigure} \\

    \begin{subfigure}{0.35\textwidth}
      \includegraphics[width=\linewidth]{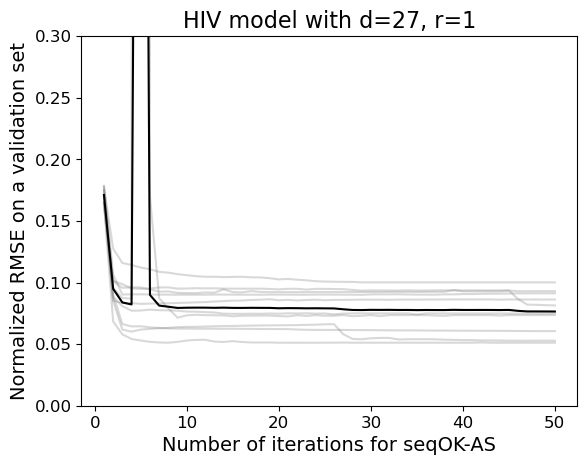}
    \end{subfigure} &
    \begin{subfigure}{0.35\textwidth}
      \includegraphics[width=\linewidth]{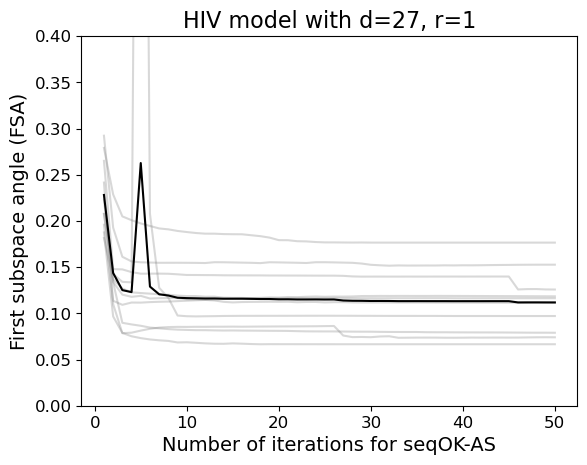}
    \end{subfigure} \\

    \begin{subfigure}{0.35\textwidth}
      \includegraphics[width=\linewidth]{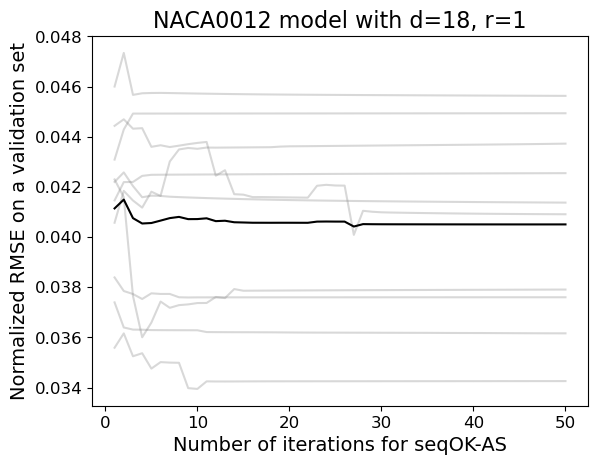}
    \end{subfigure} &
    \begin{subfigure}{0.35\textwidth}
      \includegraphics[width=\linewidth]{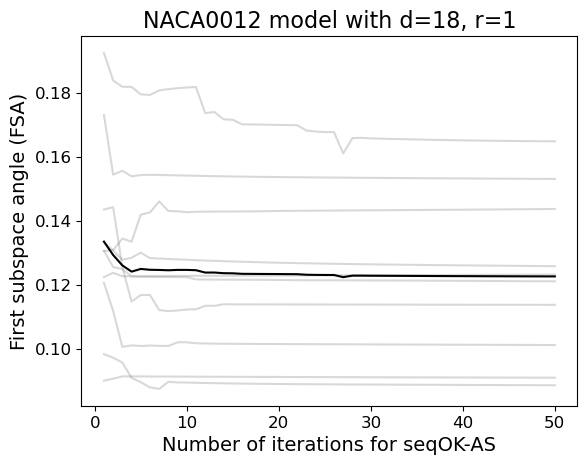}
    \end{subfigure} \\

    \begin{subfigure}{0.35\textwidth}
      \includegraphics[width=\linewidth]{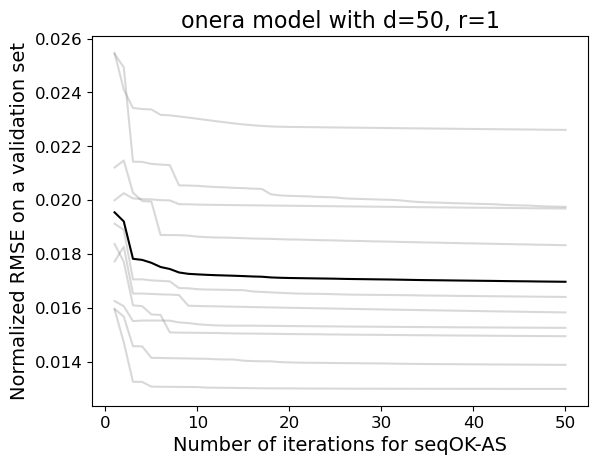}
    \end{subfigure} &
    \begin{subfigure}{0.35\textwidth}
      \includegraphics[width=\linewidth]{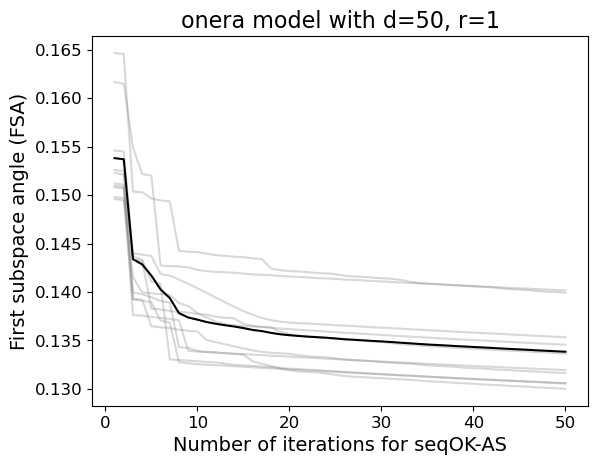}
    \end{subfigure}
  \end{tabular}
  \caption{Validation of seqOK-AS on the more realistic models (solid line: average over 10 trials)}
  \label{fig:rmse_fsa_physics_test}
\end{figure}

 For the quadratic model with effective dimensions $r=1$ and $r=2$, the experiments were conducted with input dimensions $d=25, 50, 100$ (10 repetitions for each configuration). For the more realistic test problems (i.e., NACA0012 airfoil, ONERA-M6, HIV model,
Elliptic PDE), the experiments were also conducted with 10 repetitions for each problem. Fig.  \ref{fig:rmse_fsa_quadratic_tests}-\ref{fig:rmse_fsa_physics_test} show that for almost all of the test problems, our methodology seqOK-AS (Algorithm \ref{alg:seqSAAS}) 
achieves its goals, i.e., both the normalised RMSE and first subspace angle (FSA) are substantially reduced as the number of iterations $K$ increases. In other words, the approximation accuracy of the GP posterior mean and the AS estimation are sequentially improved
using seqOK-AS. 
\par For most test problems, both RMSE and FSA decrease quickly during the first few seqOK-AS iterations, after which performance improvements are no longer visible. Nonetheless, Sec. \ref{sec:surrogate_comparison_as}-\ref{sec:method_comparison_kdr} will demonstrate 
that the contribution brought by the few successful seqOK-AS iterations is significant in comparison with other existing methods such OK-AS \cite{Palar2018}, SAAS \cite{Wycoff2021}, and GP-PLSK \cite{bouhlel_improved_2016}. Figure \ref{fig:rmse_fsa_physics_test} shows
that seqOK-AS exhibits a negative peak in performance at some iteration $k\leq 50$ for one of the 10 repetitions of the Elliptic PDE ($k=25$) and the HIV model ($k=6$), respectively. Fortunately, the undesired peak disappears quickly for both models as $k$ increases. 
The peaks are probably due to bad local minima incurred by the GP hyperparameters during training, which can also lead to poor active subspace estimation. The quick recovery is probably due to the GPs being able to achieve a good predictive performance even from a poor 
active subspace rotation, followed by a quick decay in both FSA and RMSE during the next few seqOK-AS iterations. To fully avoid the bad local minima during training, one can restart the marginal likelihood maximization algorithm a large number of times, at the cost of
increasing the training time significantly. One shortcoming to be addressed in future work is the lack of an early stopping criterion. Indeed, we can see that for the Elliptic PDE model, one would ideally stop the seqOK-AS algorithm after $k=10$ iterations, according 
to the FSA metric. Unfortunately, it is hard to test the active subspace estimation accuracy for an expensive simulator without access to true gradients.
\subsection{Comparison between seqOK-AS and other gradient-free AS methods}\label{sec:surrogate_comparison_as}
\begin{figure}[htbp]
  \centering
  \begin{tabular}{ccc}
    \begin{subfigure}{0.31\textwidth}
      \includegraphics[width=\linewidth]{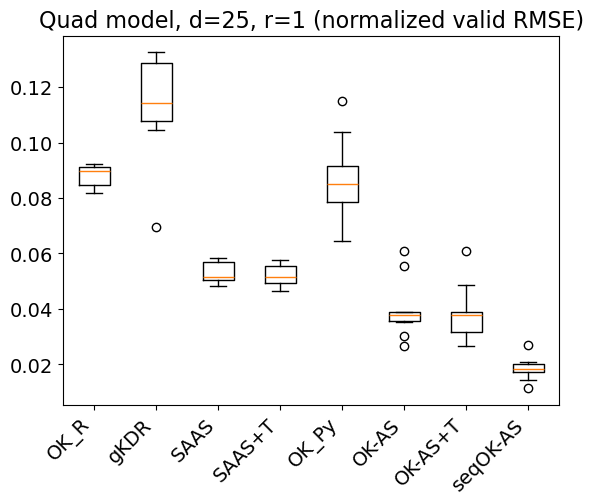}
    \end{subfigure} &
    \begin{subfigure}{0.31\textwidth}
      \includegraphics[width=\linewidth]{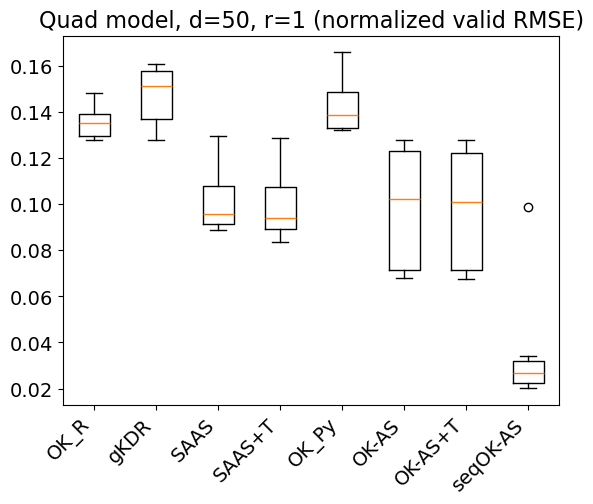}
    \end{subfigure} &
    \begin{subfigure}{0.31\textwidth}
      \includegraphics[width=\linewidth]{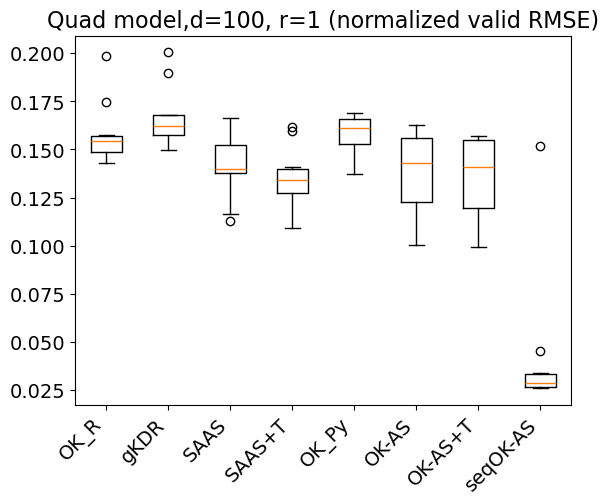}
    \end{subfigure} \\

    \begin{subfigure}{0.31\textwidth}
      \includegraphics[width=\linewidth]{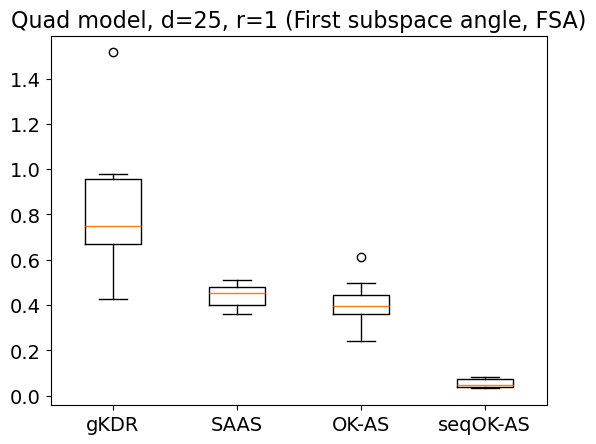}
    \end{subfigure} &
    \begin{subfigure}{0.31\textwidth}
      \includegraphics[width=\linewidth]{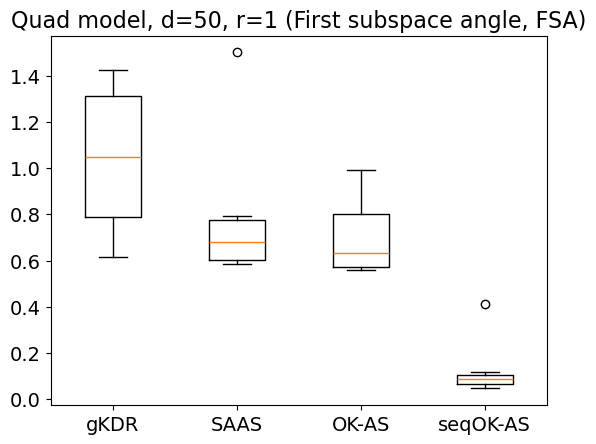}
    \end{subfigure} &
    \begin{subfigure}{0.31\textwidth}
      \includegraphics[width=\linewidth]{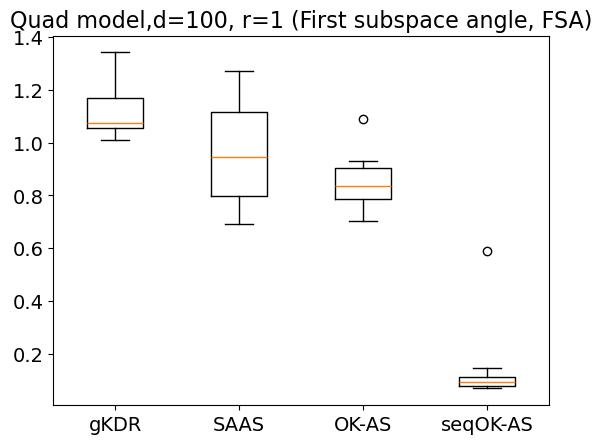}
    \end{subfigure} \\

    \begin{subfigure}{0.31\textwidth}
      \includegraphics[width=\linewidth]{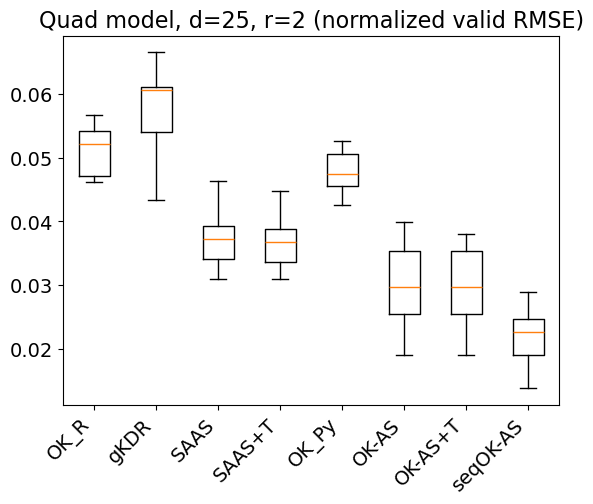}
    \end{subfigure} &
    \begin{subfigure}{0.31\textwidth}
      \includegraphics[width=\linewidth]{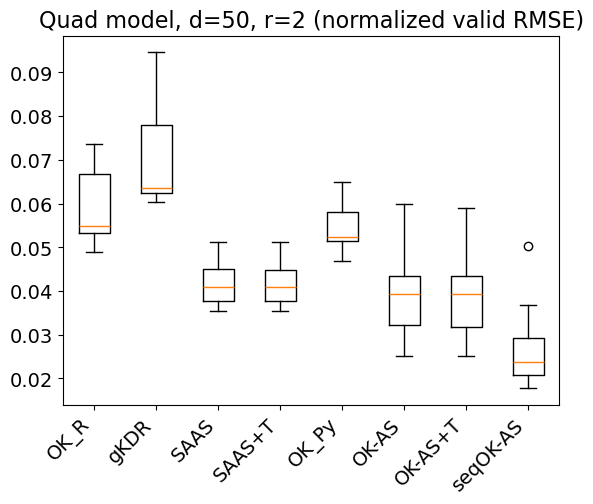}
    \end{subfigure} &
    \begin{subfigure}{0.31\textwidth}
      \includegraphics[width=\linewidth]{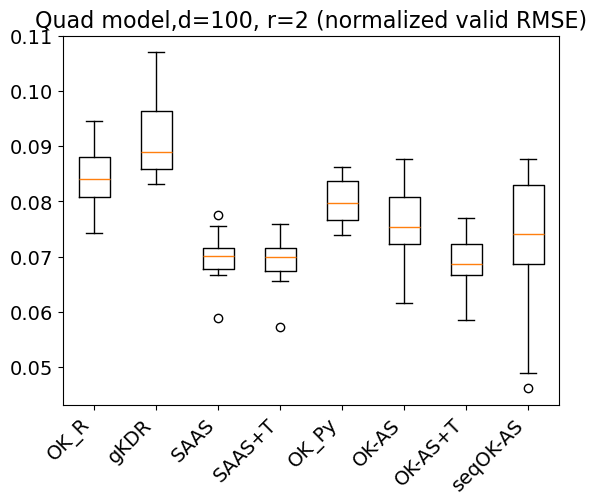}
    \end{subfigure} \\

    \begin{subfigure}{0.31\textwidth}
      \includegraphics[width=\linewidth]{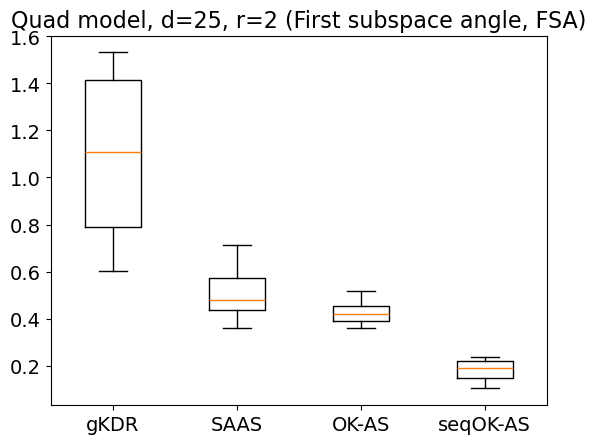}
    \end{subfigure} &
    \begin{subfigure}{0.31\textwidth}
      \includegraphics[width=\linewidth]{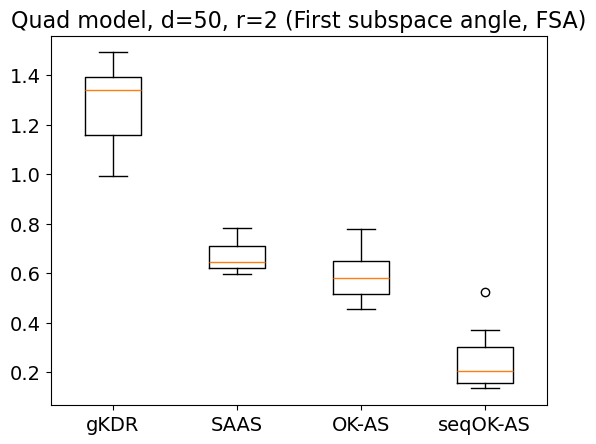}
    \end{subfigure} &
    \begin{subfigure}{0.31\textwidth}
      \includegraphics[width=\linewidth]{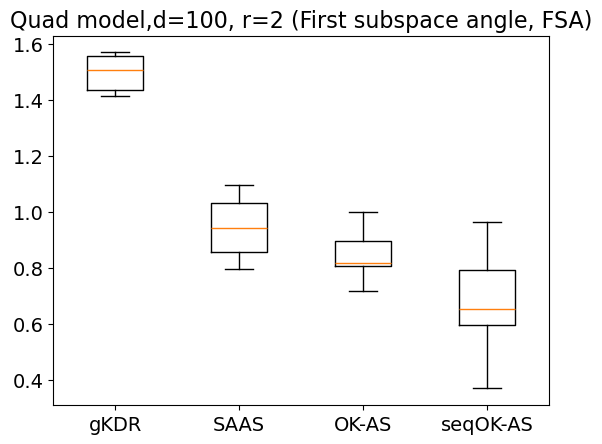}
    \end{subfigure}
  \end{tabular}
  \caption{Method comparison for GP surrogates and AS estimation for the quadratic test problems}
  \label{fig:rmse_fsa_quadratic_comparison}
\end{figure}

\begin{figure}[htbp]
  \centering
  \begin{tabular}{cc}
    \begin{subfigure}{0.35\textwidth}
      \includegraphics[width=\linewidth]{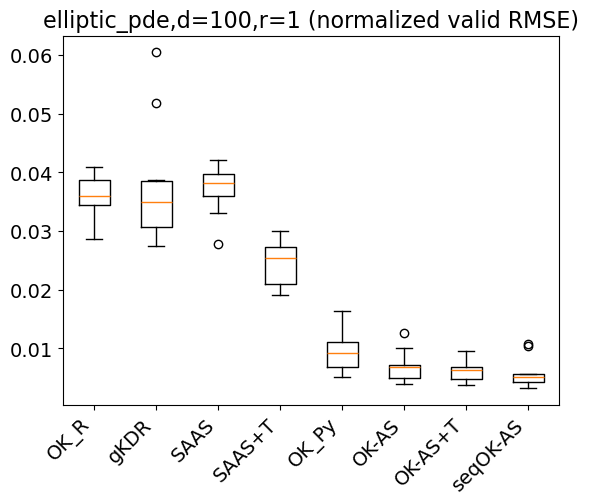}
    \end{subfigure} &
    \begin{subfigure}{0.35\textwidth}
      \includegraphics[width=\linewidth]{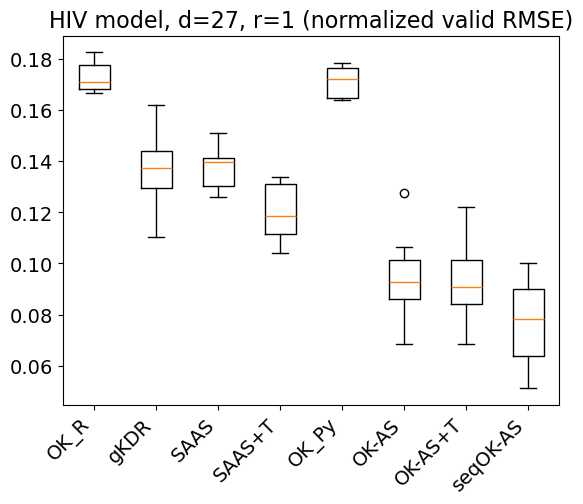}
    \end{subfigure} \\
    \begin{subfigure}{0.35\textwidth}
      \includegraphics[width=\linewidth]{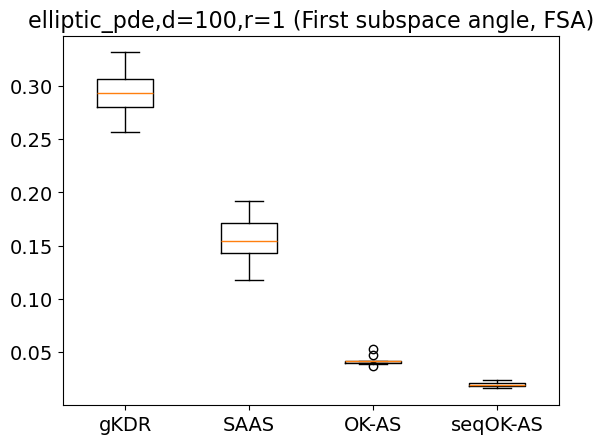}
    \end{subfigure} &
    \begin{subfigure}{0.35\textwidth}
      \includegraphics[width=\linewidth]{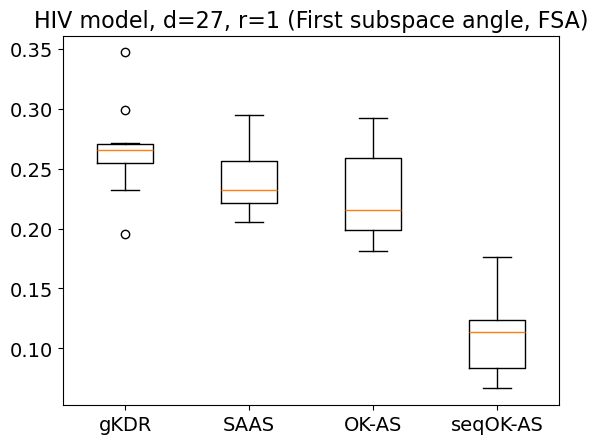}
    \end{subfigure} \\

    \begin{subfigure}{0.35\textwidth}
      \includegraphics[width=\linewidth]{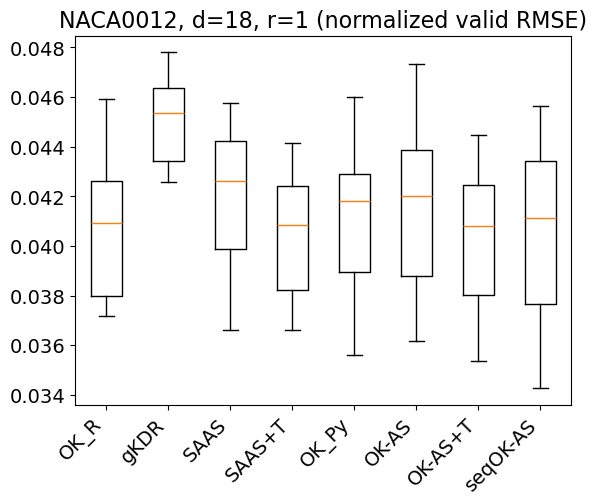}
    \end{subfigure} &
    \begin{subfigure}{0.35\textwidth}
      \includegraphics[width=\linewidth]{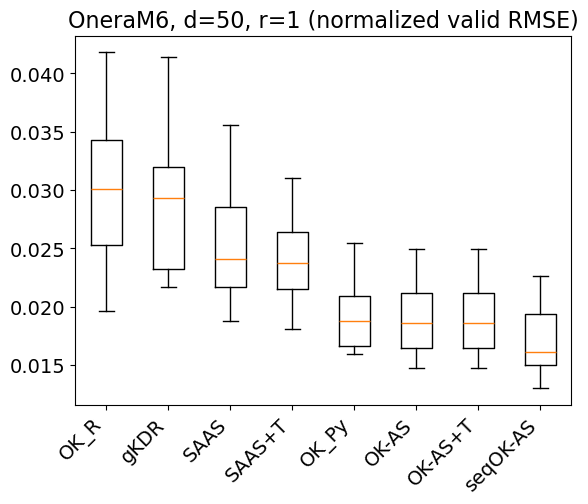}
    \end{subfigure} \\
    \begin{subfigure}{0.35\textwidth}
      \includegraphics[width=\linewidth]{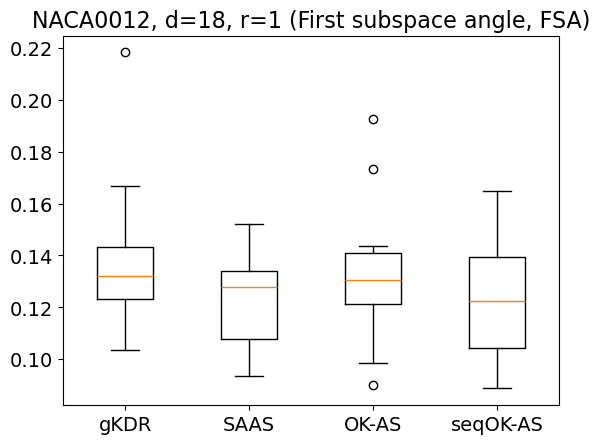}
    \end{subfigure} &
    \begin{subfigure}{0.35\textwidth}
      \includegraphics[width=\linewidth]{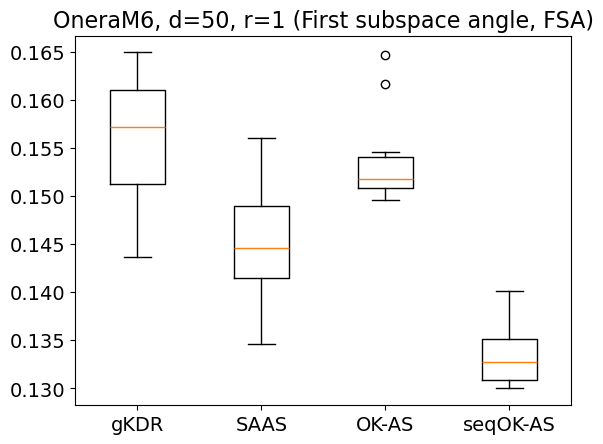}
    \end{subfigure}
  \end{tabular}
  \caption{Method comparison for GP surrogates and AS estimation for the more realistic test problems}
  \label{fig:rmse_fsa_physics_comparison}
\end{figure}

In Fig. \ref{fig:rmse_fsa_quadratic_comparison}-\ref{fig:rmse_fsa_physics_comparison} and Tab. \ref{table:rmse_physics_comparison_mean_sigma}-\ref{table:fsa_physics_comparison_mean_sigma}, the performance of seqOK-AS is compared with other existing gradient-free active subspace estimation methods and other GP regression methods from the literature:
\begin{itemize}
    \item OK\_R: high-dimensional ordinary kriging with default settings in the R package DiceKriging \cite{roustant2012dicekriging}
    \item gKDR: gradient Kernel Dimension Reduction \cite{Fukumizu2014}; the resulting estimate $\hat{\mathbf{W}}_{gKDR}\in\mathbb{R}^{d\times r}$ from (\ref{eq:gkdr_fukumizu}) is used within the training set $\{\hat{\mathbf{W}}_{gKDR}^T\mathbf{x}_i, f(\mathbf{x_i})\}_{i=1}^n$ for low-dimensional GP regression, as in \cite{Liu2016}
    \item SAAS: the AS is estimated using the Surrogate-assisted active subspace method (SAAS) \cite{Wycoff2021}; the GP surrogate is then re-trained in the SAAS rotated coordinate system, as per \cite{Wycoff2022}
    \item SAAS+T: together with the SAAS rotation, a potential truncation to dimension $d_{red}\leq d$ is applied to the training inputs \cite{Wycoff2022}; we select the low-dimensionality $d_{red}$ that minimizes RMSE
    \item OK\_{Py}: high-dimensional ordinary kriging using the Python package GPyTorch \cite{gardner2018gpytorch}
    \item OK-AS (Algorithm \ref{alg:SAAS}): the AS is estimated using the gradient of the GP posterior mean returned by OK\_{Py}; the GP surrogate is then re-trained in the OK-AS rotated coordinate system, as per \cite{Palar2018} 
    \item OK-AS+T: together with the OK-AS rotation, a potential truncation to dimension $d_{red}\leq d$ is applied to the training inputs \cite{Palar2018}; we select the low-dimensionality $d_{red}$ that minimizes RMSE
    \item seqOK-AS: our proposed method (Algorithm \ref{alg:seqSAAS} with $K=50$ iterations and no truncation, i.e., $r_k=d$ for all $k$; \ValentinCorrections{to estimate the active subspace, we select the dominant $r<d$ eigenvectors $\mathbf{W}_K[:, 1:r]$ of the final estimate of the gradient covariance matrix, $\mathbf{H}_K$).}
\end{itemize}
\par From Fig. \ref{fig:rmse_fsa_quadratic_comparison}, we draw the following conclusions. For the quadratic problems with low dimensionality $r=1$, our method seqOK-AS brings a significant improvement in terms of both RMSE and FSA compared to the existing methodologies. 
For the quadratic problems with low dimensionality $r=2$ and $d\in\{25, 50\}$, seqOK-AS significantly outperforms gKDR, SAAS and OK-AS in terms of the FSA metric; also, seqOK-AS has the lowest median RMSE among all the methods. For $d=100$ and $r=2$, seqOK-AS has the lowest
median FSA and is competitive with the other methods in terms of RMSE over the 10 repeated trials.
\par Fig. \ref{fig:rmse_fsa_physics_comparison} demonstrates the following. Regarding the more realistic test problems `Elliptic PDE model' and `HIV model', seqOK-AS is the best method in terms of the FSA metric, while also having the lowest median RMSE. For the 
physical problem `ONERA-M6 model', seqOK-AS has the lowest median RMSE and the lowest median FSA. Finally, for the physical problem `NACA0012', seqOK-AS has the lowest median FSA, while being competitive with the other approaches on the RMSE metric.

\par For completeness, Tab. \ref{table:rmse_physics_comparison_mean_sigma}-\ref{table:fsa_physics_comparison_mean_sigma} show a statistical comparison of the different methods in terms of RMSE and FSA, respectively, based on the 10 repeated trials. For most of 
the test problems, our method seqOK-AS is the sole best performing approach in terms of having the lowest average RMSE and the lowest average FSA. Table \ref{table:rmse_physics_comparison_mean_sigma} illustrates that seqOK-AS is the sole best method on the average RMSE 
metric except for the following four problems: `Quadratic model $d=100, r=2$', `Elliptic PDE model', `ONERA-M6 model', and `NACA0012 model'. Still, seqOK-AS is among the best performing methods for all of these problems. Regarding the average FSA metric, 
Table \ref{table:fsa_physics_comparison_mean_sigma} shows that seqOK-AS is the sole best performing methodology on all the test problems, except for the `NACA0012 model'. Nonetheless, seqOK-AS is among the best performing approaches even for the `NACA0012 model'. 
\begin{table}[ht]
\centering

\begin{minipage}{\textwidth}
\centering
\begin{tabular}{|l|c|c|c|c|c|c|}
\hline
\multicolumn{7}{|c|}{\textbf{Average RMSE $\pm$ one standard deviation over 10 repetitions}} \\
\hline
Test problem & d & r &OK\_R&gKDR&SAAS&SAAS+T\\
\hline
Quadratic & 25 & 1& $0.09\pm 0.004$&$0.11\pm 0.018$& $0.05\pm 0.004$& $0.05\pm 0.004$\\
Quadratic & 25 & 2 & $0.05\pm 0.004$&$0.06\pm 0.007$& $0.04\pm 0.005$& $0.04\pm 0.004$\\
Quadratic & 50 & 1 & $0.14\pm 0.006$&$0.15\pm 0.012$& $0.10\pm 0.013$& $0.10\pm 0.015$\\
Quadratic & 50 & 2 & $0.06\pm 0.009$&$0.07\pm 0.012$& $0.04\pm 0.005$& $0.04\pm 0.005$\\
Quadratic & 100 & 1 & $0.16\pm 0.016$&$0.17\pm 0.015$& $0.14\pm 0.017$& $0.13\pm 0.016$\\
Quadratic & 100 & 2 & $0.08\pm 0.006$&$0.09\pm 0.007$& $\bf{0.07\pm 0.005}$& $\bf{0.07\pm0.005}$\\
Elliptic PDE & 100 & 1 & $0.04\pm 0.004$&$0.04\pm 0.010$& $0.04\pm 0.004$& $0.02\pm 0.004$\\
HIV model & 27 & 1 & $0.17\pm 0.006$&$0.14\pm 0.014$& $0.14\pm 0.007$& $0.12\pm 0.010$\\
NACA0012 & 18 & 1 & $\bf{0.04\pm0.003}$&$0.05\pm0.002$& $\bf{0.04\pm0.003}$& $\bf{0.04\pm0.003}$\\
ONERA-M6 & 50 & 1 & $0.03\pm 0.006$&$0.03\pm 0.006$& $0.03\pm 0.005$& $0.02\pm 0.004$\\
\hline
\end{tabular}
\end{minipage}

\vspace{1em} 

\begin{minipage}{\textwidth}
\centering
\begin{tabular}{|l|c|c|c|c|c|c|}
\hline
\multicolumn{7}{|c|}{\textbf{Average RMSE $\pm$ one standard deviation over 10 repetitions}} \\
\hline
Test problem & d & r &OK\_Py&OK-AS&OK-AS+T&seqOK-AS\\
\hline
Quadratic & 25 & 1& $0.09\pm 0.015$& $0.04\pm 0.010$& $0.04\pm 0.009$& $\bf{0.02\pm 0.004}$ \\
Quadratic & 25 & 2 & $0.05\pm 0.003$& $0.03\pm 0.006$& $0.03\pm 0.006$& $\bf{0.02\pm 0.004}$\\
Quadratic & 50 & 1 & $0.14\pm 0.012$& $0.10\pm 0.024$& $0.10\pm 0.024$& $\bf{0.03\pm 0.022}$\\
Quadratic & 50 & 2 & $0.05\pm 0.005$& $0.04\pm 0.010$& $0.04\pm 0.010$& $\bf{0.03\pm 0.009}$\\
Quadratic & 100 & 1 & $0.16\pm 0.010$& $0.14\pm 0.021$& $0.14\pm 0.020$& $\bf{0.04\pm 0.037}$\\
Quadratic & 100 & 2 & $0.08\pm0.004$ & $0.08\pm0.008$ & $\bf{0.07\pm0.006}$ & $\bf{0.07\pm0.014}$\\
Elliptic PDE & 100 & 1 & $0.01\pm 0.003$& $\bf{0.007\pm 0.003}$& $\bf{0.006\pm 0.002}$& $\bf{0.006\pm 0.002}$\\
HIV model & 27 & 1 & $0.17\pm 0.006$& $0.09\pm 0.015$& $0.09\pm 0.015$& $\bf{0.08\pm 0.016}$\\
NACA0012 & 18 & 1 & $\bf{0.04\pm0.003}$& $\bf{0.04\pm0.003}$& $\bf{0.04\pm0.002}$& $\bf{0.04\pm0.004}$\\
ONERA-M6 & 50 & 1 & $0.02\pm 0.003$& $\bf{0.019\pm0.003}$& $\bf{0.019\pm0.003}$& $\bf{0.017\pm0.003}$\\
\hline
\end{tabular}
\end{minipage}
\caption{Average (normalized) Root Mean Squared Error (RMSE) with one standard deviation over 10 repeated trials. For each problem, we bold the best model and
models whose means are not statistically significantly different (significance level: $0.05$) according to a 1-sided t-test against the best model.}
\label{table:rmse_physics_comparison_mean_sigma}
\end{table}

\begin{table}[ht]
\centering

\begin{tabular}{|l|c|c|c|c|c|c|}
\hline
\multicolumn{7}{|c|}{\textbf{Average FSA $\pm$ one standard deviation over 10 repetitions}} \\
\hline
Test problem & d & r &gKDR &SAAS&OK-AS&seqOK-AS\\
\hline
Quadratic & 25 & 1& $0.83\pm 0.28$& $0.44\pm 0.05$& $0.41\pm 0.09$& $\bf0.06\pm 0.02$\\
Quadratic & 25 & 2 & $1.09\pm 0.34$ & $0.51\pm 0.10$& $0.42\pm 0.05$& $\bf{0.18\pm 0.04}$\\
Quadratic & 50 & 1 &$1.05\pm 0.29$& $0.76\pm 0.26$& $0.70\pm0.14$& $\bf{0.11\pm0.10}$\\
Quadratic & 50 & 2 &$1.28\pm 0.16$& $0.66\pm 0.06$& $0.59\pm 0.10$& $\bf{0.25\pm 0.12}$\\
Quadratic & 100 & 1 &$1.12\pm 0.10$& $0.95\pm 0.19$& $0.84\pm 0.11$& $\bf{0.14\pm 0.15}$\\
Quadratic & 100 & 2 &$1.50\pm 0.06$& $0.94\pm 0.10$& $0.84\pm 0.09$& $\bf{0.66\pm0.18}$\\
Elliptic PDE & 100 & 1 & $0.29\pm 0.02$ & $0.15\pm 0.02$& $0.04\pm 0.004$& $\bf{0.02\pm 0.002}$\\
HIV model & 27 & 1 &$0.27\pm 0.04$& $0.24\pm 0.03$& $0.23\pm 0.04$& $\bf{0.11\pm 0.03}$\\
NACA0012 & 18 & 1 &$\bf{0.14\pm 0.03}$& $\bf{0.12\pm 0.02}$& $\bf{0.13\pm 0.03}$& $\bf{0.12\pm 0.02}$\\
ONERA-M6 & 50 & 1 &$0.16\pm 0.006$& $0.15\pm 0.006$& $0.15\pm 0.005$& $\bf{0.13\pm 0.004}$\\
\hline
\end{tabular}

\caption{Average First Subspace Angle (FSA) with one standard deviation over 10 repeated trials. For each problem, we bold the best model and
models whose means are not statistically significantly different (significance level: $0.05$) according to a 1-sided t-test against the best model.}
\label{table:fsa_physics_comparison_mean_sigma}
\end{table}
\subsection{Further comparison between seqOK-AS and other regression methods}\label{sec:method_comparison_kdr}
One of the main aims of seqOK-AS is to provide an accurate gradient-free active subspace estimation.
We have already shown that seqOK-AS outperforms others active subspace methods such as g-KDR, OK-AS and SAAS on most of the test problems from Sec. \ref{sec:surrogate_comparison_as}. This section explores whether seqOK-AS is 
competitive with other state-of-the-art surrogate methods on the same test problems in terms of predictive performance (i.e., validation RMSE). Most of the surrogate methods considered are discussed in Section \ref{sec:dim_red_summary}: 
GP-KDR (\ref{eq:gp_kdr}), GP-EL, GP-PLS, and GP-PLSK. \ValentinCorrections{Polynomial Chaos Expansion (PCE, \cite{ghanem_stochastic_1991}) is a popular surrogate method that is included here for predictive comparison.} 
The active subspace estimation accuracy is not addressed in this section, since KDR, EL, and PLS are not AS estimation methods. 
\par KDR is implemented as in \cite{dimred_kdr}. Once the KDR estimator has been obtained, the GP-KDR surrogate (\ref{eq:gp_kdr}) is trained using the Surrogate Modeling Toolbox (SMT) \cite{SMT2019}. 
GP-EL is implemented using the GPyTorch library \cite{gardner2018gpytorch}, and it is trained using Adam (Adaptive Moment Estimation) gradient descent \cite{Kingma2014AdamAM}. GP-PLS and GP-PLSK are implemented and trained using SMT \cite{SMT2019}. \ValentinCorrections{PCE is constructed and trained using the UQLab toolbox \cite{UQdoc_22_104} via the Least Angle Regression (LAR) strategy \cite{Blatman2011}.
LAR is a commonly used sparse regression procedure for training PCE surrogates in high-dimensions.}
\par
\begin{figure}[htbp]
  \centering
  \begin{tabular}{ccc}
    \begin{subfigure}{0.31\textwidth}
      \includegraphics[width=\linewidth]{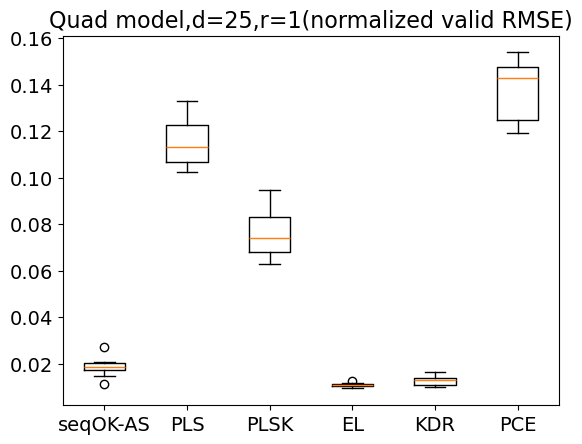}
    \end{subfigure} &
    \begin{subfigure}{0.31\textwidth}
      \includegraphics[width=\linewidth]{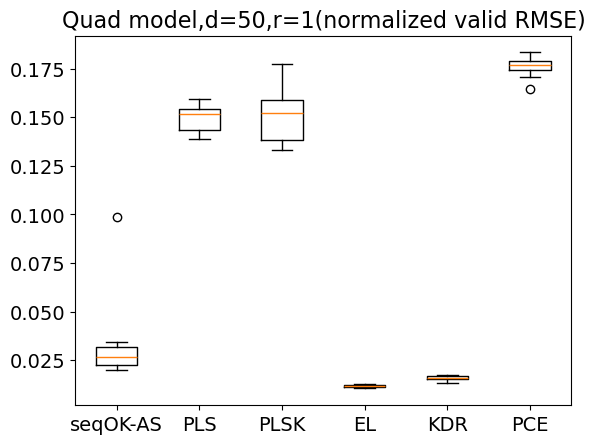}
    \end{subfigure} &
    \begin{subfigure}{0.31\textwidth}
      \includegraphics[width=\linewidth]{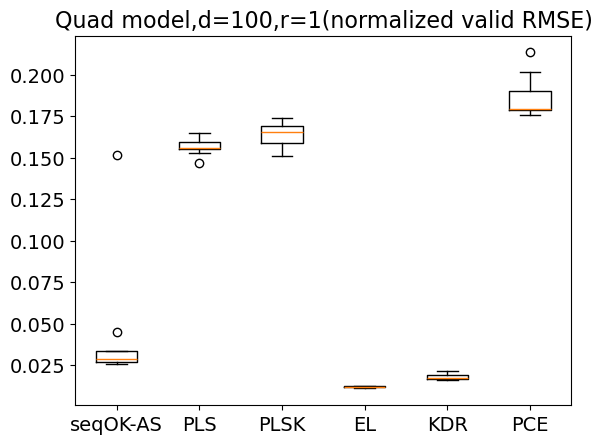}
    \end{subfigure} \\

    \begin{subfigure}{0.31\textwidth}
      \includegraphics[width=\linewidth]{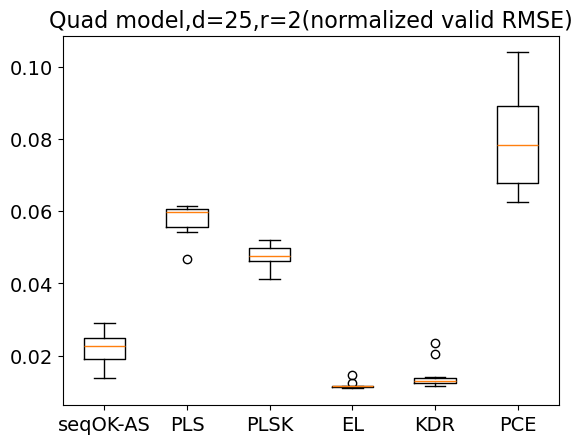}
    \end{subfigure} &
    \begin{subfigure}{0.31\textwidth}
      \includegraphics[width=\linewidth]{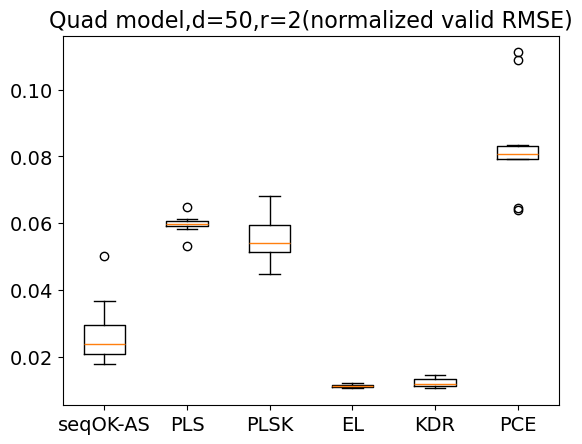}
    \end{subfigure} &
    \begin{subfigure}{0.31\textwidth}
      \includegraphics[width=\linewidth]{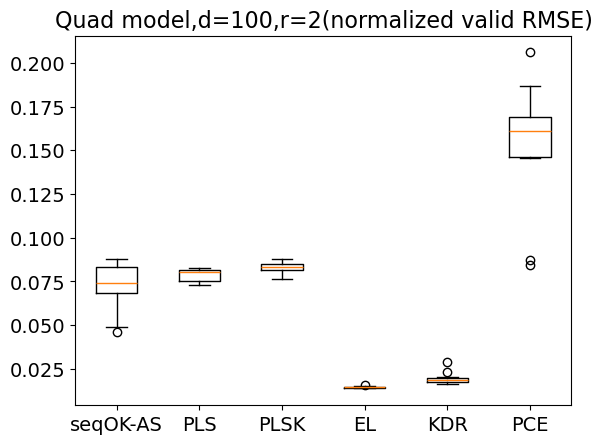}
    \end{subfigure} \\

  \end{tabular}
  \caption{GPR \ValentinCorrections{and PCE} comparison for the quadratic test problems (seqOK-AS vs popular methods)}
  \label{fig:rmse_fsa_quadratic_comparison_short}
\end{figure}

\begin{figure}[htbp]
  \centering
  \begin{tabular}{cc}
    \begin{subfigure}{0.4\textwidth}
      \includegraphics[width=\linewidth]{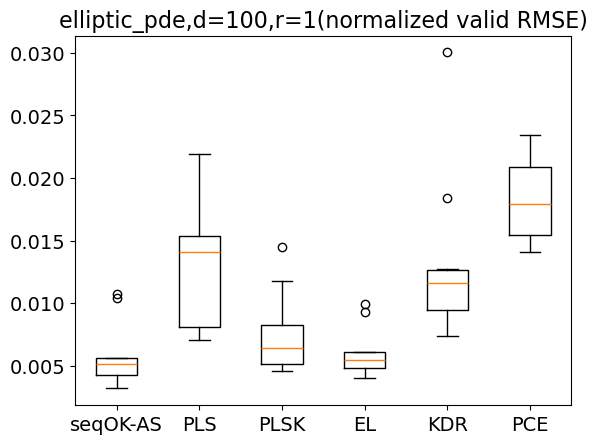}
    \end{subfigure} &

    \begin{subfigure}{0.4\textwidth}
      \includegraphics[width=\linewidth]{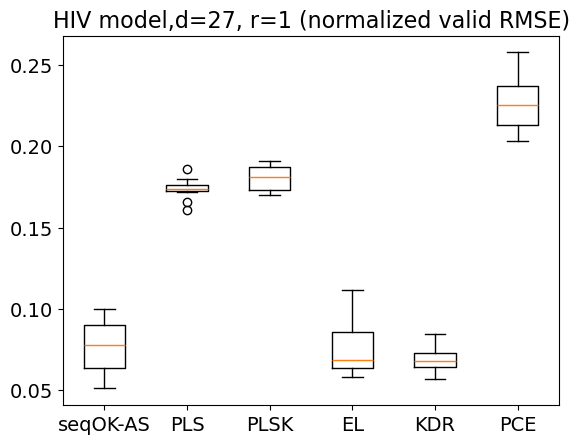}
    \end{subfigure} \\

    \begin{subfigure}{0.4\textwidth}
      \includegraphics[width=\linewidth]{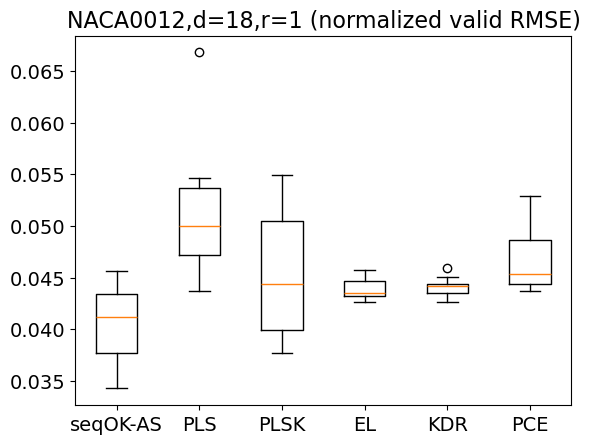}
    \end{subfigure} &

    \begin{subfigure}{0.4\textwidth}
      \includegraphics[width=\linewidth]{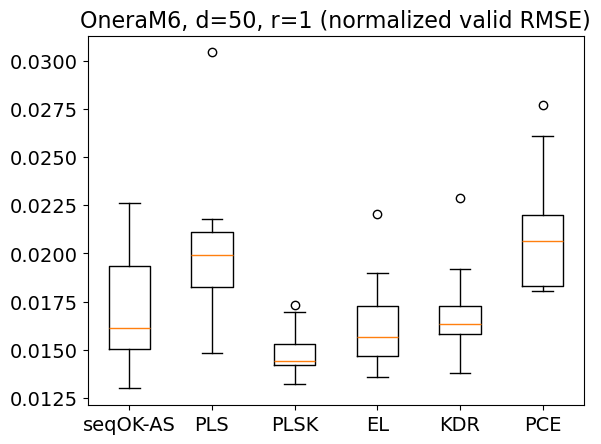}
    \end{subfigure}
  \end{tabular}
  \caption{GPR \ValentinCorrections{and PCE} comparison for the more realistic problems (seqOK-AS vs popular methods)}
  \label{fig:rmse_fsa_physics_comparison_short}
\end{figure}

 On one hand, Fig. \ref{fig:rmse_fsa_quadratic_comparison_short} shows that for the quadratic test problems, our method seqOK-AS significantly outperforms GP-PLS, GP-PLSK \ValentinCorrections{and PCE} in almost all cases, except for $d=100$ and $r=2$, where seqOK-AS still has a lower median RMSE.
On the other hand, KDR and EL demonstrate a small, consistent advantage versus seqOK-AS on problems with $r=1$. Further, the advantage of KDR and EL over seqOK-AS grows for the quadratic problems with $r=2$. Regarding the more realistic test problems,
Fig. \ref{fig:rmse_fsa_physics_comparison_short} shows that seqOK-AS is among the best-performing methods in all test cases. In particular, seqOK-AS has the lowest median validation error among all methods for the `NACA0012 model'. 
Note that GP-KDR (\ref{eq:gp_kdr}) and GP-EL perform similarly in all the test problems; this was to be expected, since the two dimension reduction methods are nearly identical if a linear kernel for $K_Y$ is used for KDR in (\ref{eq:kdr_fukumizu}). 
In our experiments, we use an RBF kernel for $K_Y$ of KDR, in line with the existing literature. GP-PLSK performs similarly or better compared to GP-PLS, as also observed in \cite{bouhlel_improved_2016}. The performance of PCE might be improved via a linear or non-linear version of the PLS dimension reduction algorithm \cite{papaioannou2019pls}.
Also, the performance of GP-PLS and GP-PLSK might be further improved by adopting a nonlinear PLS version \cite{Ehre2022}. 
\par To the best of our knowledge, all these GP surrogate options for ridge approximation problems (i.e., GP-gKDR, OK-AS, SAAS, GP-KDR, GP-EL, GP-PLS, and GP-PLSK) have never been compared on a benchmark of test cases. 
Thus, the method comparison presented in Sec. \ref{sec:surrogate_comparison_as}-\ref{sec:method_comparison_kdr} is another novel contribution of the current paper, in addition to introducing the new approach seqOK-AS. 
Our method comparison is inspired by the source of our test problems \cite{Gautier2022}, where two of the above methods (i.e., SAAS and GP-EL) were benchmarked against each other.

\section{Numerical results for rare event probability estimation}\label{sec:method_comparisonzz}
In Section \ref{sec:ce_seqsaas}, we introduced a new approach \textit{iCEred+seqOK-AS}. Our method is based on the pre-existing Cross Entropy methodology of \cite{Uribe2021cross} and the new GP surrogate method seqOK-AS (Algorithm \ref{alg:seqSAAS}). Recall that seqOK-AS was validated in the context of GP regression and active subspace estimation in Section \ref{sec:numerical_results}.
\par Firstly, we validate iCEred+seqOK-AS on two reliability test problems. The first test case is a quadratic example in dimension $d=100$: $\mathbf{X}$ is a multivariate standard normal, $f(\mathbf{x})=-4-\frac{5}4(x_1-x_2)^2+\frac{1}{\sqrt{d}}\sum_{i=1}^d x_i$, featuring an active subspace of dimension $r=2$ and associated with a failure probability $P_f=6.62\cdot 10^{-6}$ (\ref{eq:MC_rare_events}). A second test case deals with the failure of a 23-bar truss assembly, from \cite{Lee2006}, in dimension $d=10$, with an active subspace of dimension $r=1$ and $P_f=1.02\cdot 10^{-8}$. For the second test case, see \cite{Lee2006} for the physical interpretation and probabilistic description of the 10D random vector $\mathbf{X}$. 
\ValentinCorrections{Each test problem has a unique design point, i.e., a unique failure mode.  On these two test cases, we compare our approach iCEred+seqOK-AS versus two classical reliability methods: FORM and Importance Sampling using FORM initialization (FORM-IS) \cite{schueller_critical_1987}. To analyse the impact of our surrogate method seqOK-AS versus the existing approaches OK\_Py and OK-AS, we created two baseline versions iCEred+OK\_Py and iCEred+OK-AS to be compared with iCEred+seqOK-AS; recall that OK\_Py and OK-AS were introduced for the regression benchmark in Sec. \ref{sec:surrogate_comparison_as}. To clarify, iCEred+OK\_Py uses the gradient of the posterior mean of a standard high-dimensional GP metamodel $\tilde{m}_{j}(\mathbf{x})\approx L^{\varepsilon_j}(\mathbf{x})$ to estimate $\nabla L^{\varepsilon_j}(\mathbf{x})$ for iCEred (\ref{eq:Fisher_info_generic}), while iCEred+OK-AS uses one rotation to build an OK-AS metamodel for $L^{\varepsilon_j}(\mathbf{x})$, followed by estimating $\nabla L^{\varepsilon_j}(\mathbf{x})$.} In addition, we consider two other
competitive gradient-free methods, since these approaches have already demonstrated good performance on these two reliability problems \cite{Ehre2022}. The two methods are known as Sequential Importance Sampling (SIS) \cite{Papaioannou2016sequential} and Active Sequential Subspace Importance Sampling (ASSIS) \cite{Ehre2022}.\\
\noindent\textbf{Sequential Importance Sampling (SIS).} Sequential Importance Sampling is a standard approach for reliability in high-dimensional problems; it does not involve a metamodel, nor dimension reduction. \ValentinCorrections{Nonetheless, the SIS method is the source of inspiration for the sequential algorithm iCEred described in Sec. \ref{sec:ce_seqsaas}. To estimate the failure probability, SIS uses a sequence of smoothed biasing densities of the form (\ref{eq:smoothing_is}), parametrized by decreasing temperatures $\infty=\varepsilon_0\geq\varepsilon_1\geq \dots\geq\varepsilon_m \approx 0$. For this purpose, a collection of $N_{SIS}$ particles starts from $q$ ($\varepsilon_0\approx\infty$) and then sequentially approaches an approximation of $q^*_{IS}$ (\ref{eq:optimal_is_density}) ($\varepsilon_m \approx 0$). The particles move from one temperature to the next (i.e., from $\pi^{\varepsilon_{j-1}}(\mathbf{x})\propto h^{\varepsilon_{j-1}}(-f(\bs{x}))q(\bs{x})$ to $\pi^{\varepsilon_{j}}(\mathbf{x})\propto h^{\varepsilon_{j}}(-f(\bs{x}))q(\bs{x})$) using Markov Chains, as common in Sequential Monte Carlo procedures \cite{Del2006sequential}. The criterion for each transition step $\varepsilon_{j-1}\to\varepsilon_{j}$ is very similar to the iCEred+seqOK-AS procedure in Sec. \ref{sec:ce_seqsaas}, i.e., a constant coefficient of variation is enforced:
 $$\varepsilon_j=\argminE_{\varepsilon\in(0, \varepsilon_{j-1})}\Bigg(\widehat{\text{CoV}}\Bigg(\Bigg\{\frac{h^{\varepsilon}(-f(\bs{x}_i^{(j-1)}))q(\bs{x}_i^{(j-1)})}{h^{\varepsilon_{j-1}}(-f(\bs{x}_i^{(j-1)}))q(\bs{x}_i^{(j-1)})}\Bigg\}_{i=1}^{N_{SIS}}\Bigg)-\delta_{SIS}\Bigg)^2=\argminE_{\varepsilon\in(0, \varepsilon_{j-1})}\Bigg(\widehat{\text{CoV}}\Bigg(\Bigg\{\frac{h^{\varepsilon}(-f(\bs{x}_i^{(j-1)}))}{h^{\varepsilon_{j-1}}(-f(\bs{x}_i^{(j-1)}))}\Bigg\}_{i=1}^{N_{SIS}}\Bigg)-\delta_{SIS}\Bigg)^2.$$
Furthermore, the stopping criterion for SIS is identical to the iCEred+seqOK-AS stopping criterion (\ref{eq:stopping_criterion_is}).}\\
\noindent\textbf{Active Sequential Subspace Importance Sampling (ASSIS).} \ValentinCorrections{The ASSIS method is motivated as an extension of the SIS method for high-dimensional problems with expensive simulators. ASSIS proposes a strategy for carefully replacing the physical model $f(\mathbf{x})$ with a surrogate approximation $\hat{f}(\mathbf{x})$ inside the SIS procedure described above. In this way, a much larger number of particles $N_{ASSIS}>N_{SIS}$ can be potentially used for estimating the probability of failure. The surrogate model is a Polynomial Chaos Expansion (PCE) coupled with dimension reduction via non-linear partial least squares. The following is a succinct summary of the ASSIS procedure. For full details, see the original source \cite{Ehre2022}.}
\par \ValentinCorrections{Similar to SIS, ASSIS works using a sequence of smoothed biasing densities, parametrized by decreasing temperatures $\infty=\varepsilon_0\geq\varepsilon_1\geq \dots\geq\varepsilon_m \approx 0$; the ASSIS particles are distributed according to $\hat{\pi}^{\varepsilon_{j}}(\mathbf{x})\propto h^{\varepsilon_{j}}(-\hat{f}(\bs{x}))q(\bs{x})$. For each $\varepsilon_j$, the surrogate accuracy $\hat{f}(\bs{x})\approx f(\bs{x})$ for the particles $\{\bs{x}_i^{(j)}\}_{i=1}^{N_{ASSIS}}$ is ensured through two strategies for constructing the training set. First, $N_{train}^{(j)}$ particles $\{\bs{x}_{k_i}^{(j)}\}_{i=1}^{N_{train}^{(j)}}$ are selected at random for model evaluation. The resulting set $\{\bs{x}_{k_i}^{(j)}, f(\bs{x}_{k_i}^{(j)})\}_{i=1}^{N_{train}^{(j)}}$ is incorporated into the PCE training set. Second, ASSIS improves surrogate accuracy through active learning by iteratively selecting particles with the highest PCE predictive uncertainty and adding them, along with their simulator evaluations, to the training set until all uncertainties fall below a threshold. The active learning step is performed in low-dimensions, with the help of the non-linear PLS dimension reduction. For this purpose, \cite{Ehre2022} proposes a new method for estimating predictive uncertainties for PCE surrogates coupled with non-linear PLS dimension reduction. Finally, the criterion for choosing $\varepsilon_{j}$ and the stopping criterion for ASSIS are similar to SIS, except that $f(\mathbf{x})$ is replaced by the surrogate $\hat{f}(\mathbf{x})$.}\\
\noindent\textbf{Choice of parameters for the method comparison.} \ValentinCorrections{SIS uses the same set of parameters for both test problems (i.e., `quadratic' and `2D truss'). That is, the number of particles is $N_{SIS}=2\cdot 10^3$, while the threshold for both the transition steps and the stopping criterion is $\delta_{SIS}=1.5$. ASSIS also uses an identical set of parameters for both example problems. Namely, the number of particles is $N_{ASSIS}=10^4$, while the threshold is still $\delta_{ASSIS}=1.5$. For each temperature $\varepsilon_j$, $N_{train}^{(j)}=5d$ particles are initially added to the ASSIS surrogate training set, after which the active learning strategy is applied. To clarify, these parameters correspond to those used in the original study \cite{Ehre2022} for the two test problems, where additional key aspects, such as the properties of the Markov chains for SIS and the configuration of the PCE surrogates with dimension reduction in ASSIS, are also discussed in detail. For the proposed iCEred+seqOK-AS approach, $N=100$ particles are used for each test problem, with thresholds $\delta=1.5$ and $\varepsilon_{AS}=0.01$ as detailed in Sec. \ref{sec:ce_seqsaas}. An additional cost of $N_{init}=300$ model evaluations is incurred for the iCEred+seqOK-AS procedure in order to warm-start the seqOK-AS surrogates, as discussed in Sec. \ref{sec:ce_seqsaas}. \SecondRoundValentinCorrections{Both parameters $N$ and $N_{init}$ are selected using trial-and-error. First, the warm-up training size $N_{init}$ is chosen such that the cross-validation error for a seqOK-AS surrogate of $f$ is below a threshold. Second, the number of particles $N$ is selected such that the resulting scheme iCEred+seqOK-AS shows a stable convergence in a large number of repeated trials. An automatic tuning procedure is left for future work. } 
The refinement step $N_{IS}>N$ is employed only for the `2D truss' problem, using $\overline{\delta}=0.2$ for the stopping criterion in (\ref{eq:stop_criterion_refinement_IS}). For a fair comparison, the same settings of iCEred+seqOK-AS are used for the baseline versions iCEred+OK\_Py and iCEred+OK-AS. To obtain the design point, FORM uses the classical gradient-free optimisation method Sequential Least Squares Programming (SLSQP \cite{Nocedal2006}). FORM-IS uses the same design point, together with 2500 extra samples for importance sampling.}
\par For the comparisons, we conduct the reliability experiments with 100 repetitions and use four metrics:
\begin{itemize}
    \item Average number of simulator evaluations $\pm$ 2 standard deviations,
    \item Relative bias: $RB:=|P_f-\E[\hat{P}_f]|/P_f$,
    \item Coefficient of variation: $CoV:=\sqrt{\text{Var}[\hat{P}_f]}/\E[\hat{P}_f]$,
    \item Relative root mean squared error (RMSE): $\sqrt{RB^2 +CoV^2(\E[\hat{P}_f]/P_f)^2}$.
\end{itemize}
\par The performance of all the reliability methods on the two test problems is presented in Table \ref{tab:Quad_Treillis2D}. As explained above, the methods SIS, ASSIS, iCEred+OK\_Py, iCEred+OK-AS, and iCEred+seqOK-AS follow the idea of IS via smoothing (\ref{eq:smoothing_is}), and use a standard logistic function as a smooth approximation for the Heaviside function:
$$h^\varepsilon(-f(\mathbf{x}))=\frac{1}{2}\Bigg[1+\tanh\Bigg(-\frac{f(\mathbf{x})}{\varepsilon}\Bigg)\Bigg].$$
\ValentinCorrections{The logistic function is preferred to other smoothing options such as the standard Gaussian CDF or the square of the standard Gaussian CDF (\ref{eq:h_CDF}), since it is the only option that leads to a numerically stable performance during all 100 repetitions. Recall that during each iCEred iteration, the number of particles $N$ used for iCEred+seqOK-AS is low ($N=100$). Therefore, our finding is in line with the conclusions of \cite{Uribe2021cross}, which reports that for the numerical experiments with $N=100$ samples for each iCEred iteration, the logistic function appears to be more robust compared to the standard Gaussian CDF.}
\par The results for SIS and ASSIS are taken directly from \cite{Ehre2022}. \ValentinCorrections{Table \ref{tab:Quad_Treillis2D} shows that our method iCEred+seqOK-AS is the best method \ValentinCorrections{on the relative bias metric} for the quadratic reliability problem, while also being competitive in terms of cost (i.e., the average cost is second best to FORM). 
Furthermore, iCEred+seqOK-AS is the best approach for the `2D truss' reliability problem in terms of the relative bias accuracy metric, while also outperforming SIS and FORM-IS on the cost and RRMSE metrics. On both test problems, iCEred+seqOK-AS is more robust compared to the baseline versions iCEred+OK\_Py and iCEred+OK-AS, since it has a lower average cost and a lower cost dispersion (standard deviation).}
SIS exhibits a relatively large RRMSE on the `2D truss' problem due to a large CoV.
\begin{table}[H]
\centering
\caption{Results for the quadratic and 2D truss reliability problems over 100 repeated trials}
\label{tab:Quad_Treillis2D}
\begin{tabular}{ |p{2.6cm}||p{3.5cm}|p{3.1cm}|p{1.7cm}|p{2cm}|p{1.3cm}|  }
 \hline
 \multicolumn{6}{|c|}{Rare event probability estimation} \\
 \hline
 Model & Method & Cost & $\E[\hat{P}_f]$& Rel. bias & RRMSE\\
 \hline
 \multirow{7}{*}{\shortstack{Quadratic \\ ($P_f=6.62\cdot10^{-6}$)}} &  \ValentinCorrections{FORM} & \bfseries 202 & $3.17\cdot10^{-5}$ &3.781 & -\\
 & \ValentinCorrections{FORM-IS} & 2702 &$6.53\cdot10^{-6}$& 0.013 & \bfseries 0.09\\
& \ValentinCorrections{iCEred+OK\_Py} & 1034$\pm$564 &$6.80\cdot10^{-6}$& 0.028 & 0.56\\
& \ValentinCorrections{iCEred+OK-AS} & 1036$\pm$859 &$6.59\cdot10^{-6}$& 0.004 & 0.25\\
 & iCEred+seqOK-AS & 978$\pm$201 &$\mathbf{6.60\cdot10^{-6}}$& \bfseries 0.003 & 0.18\\
 & SIS & 24000$\pm$5000 &$6.57\cdot10^{-6}$& 0.008 & 0.34\\
 & ASSIS & 1900$\pm$300 & $6.34\cdot10^{-6}$&0.043 & 0.27\\
 \hline
 \multirow{7}{*}{\shortstack{2D truss \\ ($P_f=1.02\cdot10^{-8}$)}} & \ValentinCorrections{FORM} & \bfseries 165 & $2.86\cdot10^{-9}$&0.720 & -\\
 & \ValentinCorrections{FORM-IS} & 2665 & $1.08\cdot10^{-8}$& 0.060 & 0.15\\
 & \ValentinCorrections{iCEred+OK\_Py} & 1969$\pm$3864 &$1.10\cdot10^{-8}$& 0.074 & 0.16\\
& \ValentinCorrections{iCEred+OK-AS} & 1819$\pm$1269 
&$1.05\cdot10^{-8}$& 0.030 & \bfseries 0.13\\
 & iCEred+seqOK-AS & 1760$\pm$675 &$\mathbf{1.01\cdot10^{-8}}$& \bfseries 0.005 & 0.14\\
 & SIS & 25000$\pm$1000 & $9.93\cdot10^{-9}$&0.026 & 1.11\\
 & ASSIS & 350$\pm$65 & $7.44\cdot10^{-9}$&0.270 & 0.28\\ 
 \hline
\end{tabular}
\end{table}
\par Secondly, a more difficult application case deals with the 2D modeling of a steel plate whose simulator output involves a set of elliptic partial differential equations (i.e., Cauchy-Navier equations). 
The 2D plate failure problem is in dimension $d=869$ and uses a Karhunen-Loeve discretization of a log-Gaussian field for the random vector $\mathbf{X}$ (i.e, Young's modulus) \cite{Uribe2021cross, Ehre2022}. 
The results in Table \ref{tab:Acier2D} compare iCEred+seqOK-AS to SIS and FORM. \ValentinCorrections{For iCEred+seqOK-AS, we use $N=500$ particles with thresholds $\delta=1.5$ and $\varepsilon_{AS}=0.01$ as detailed in Sec. \ref{sec:ce_seqsaas}. There is no refinement in the final IS step (\ref{eq:final_is_step}), i.e., the simple option $N_{IS}=N$ is used. Further, as opposed to the `quadratic' and `2D truss' problems, we did not use any data to warm-start the seqOK-AS surrogates before running the iCEred+seqOK-AS procedure.
For SIS, we use $N_{SIS}=2\cdot 10^3$ particles, while the threshold for both the transition steps and the stopping criterion is $\delta_{SIS}=1.5$. Note that the same threshold $\delta=1.5$ is used for both iCEred+seqOK-AS and SIS. 
Despite trying to use the same modeling assumptions and numerical implementation for the PDE solver, we were unable to reproduce the results from \cite{Uribe2021cross, Ehre2022}, hence we do not have access to the true $P_f$ in this case. The costs of running a full Monte Carlo simulation would be prohibitive.}
However, the agreement between all the methods in Tab. \ref{tab:Acier2D} regarding the order of magnitude ($10^{-8}$) is encouraging. \SecondRoundValentinCorrections{The discrepancy from the order of magnitude $(10^{-6})$ reported in the literature \cite{Uribe2021cross, Ehre2022} is perhaps due to a subtle difference in the implementation of the simulator. Over the 15 repeated
trials presented in Tab. \ref{tab:Acier2D}, there was no indication that the reliability methods have failed to converge.}
\par The FORM algorithm converged using the SLSQP optimization algorithm after a large number of simulations (9571), \ValentinCorrections{given the challenge posed by the high-dimensionality $d=869$.} Our method iCEred+seqOK-AS returns the lowest cost and the lowest CoV. For completeness, we outline the findings of \cite{Ehre2022} on the steel plate problem, where a true $P_f=4.23\cdot 10^{-6}$ is reported. ASSIS returned an average cost of 1318 simulations over 100 repetitions, with a CoV of 0.021. 
In contrast, SIS had an average cost of 17000 simulations, with a CoV of 0.625; this is similar to our findings in Table \ref{tab:Acier2D}.
\par Overall, the results presented in Tab. \ref{tab:Quad_Treillis2D}-\ref{tab:Acier2D} demonstrate that the proposed approach iCEred+seqOK-AS is very competitive, offering high precision for a reduced computational cost compared to standard reliability methods, especially for non-linear or/and high-dimensional problems.
\begin{table}[H]
\centering
\caption{Results for the 2D steel plate model over 15 repeated trials}
\label{tab:Acier2D}
\begin{tabular}{ |p{4cm}||p{3cm}|p{2cm}|p{1.5cm}|  }
 \hline
 \multicolumn{4}{|c|}{Rare event probability estimation} \\
 \hline
 Method & Cost & $\E[\hat{P}_f]$ & CoV\\
 \hline
 FORM & 9571 & $6.7\cdot 10^{-8}$& -\\
 iCEred+seqOK-AS & \bfseries 3281$\pm$248 & $3.5\cdot 10^{-8}$ & \bfseries 0.09\\
 SIS & 14000$\pm$0 & $4.4\cdot 10^{-8}$ & 0.68\\
 \hline
\end{tabular}
\end{table}

\section{Conclusions and future work}
In Section \ref{sec:dim_red_gp}, we proposed seqOK-AS, a method which can significantly improve gradient-free active subspace estimation and kriging accuracy in high dimensions when compared to other popular methods such as SAAS \cite{Wycoff2022} and OK-AS \cite{Palar2018}; this is demonstrated by the numerical results in Section \ref{sec:surrogate_comparison_as}. We further demonstrate that seqOK-AS is a competitive option even when compared to other state-of-the-art high-dimensional kriging methods such as GP-KDR or GP-PLSK (Section \ref{sec:method_comparison_kdr}). In addition, our approach allows for goal-oriented dimension reduction without gradients in the context of rare event estimation (Section \ref{sec:ce_seqsaas}). The resulting rare event oriented strategy proves to be numerically efficient and robust (Section \ref{sec:method_comparisonzz}). Current limitations arise in the lack of numerical experiments for scenarios where the effective dimension is relatively high ($r \ge 5$), as well as in achieving a fair comparison of training times across the different methods. In particular, the methods rely on distinct optimization strategies and software implementations, which complicates an equitable assessment of their respective training durations. Nonetheless, in many practical applications, the training times are insignificant compared to the cost of running the numerical model e.g., $n=5d$ times for obtaining a training set. Ongoing work focuses on applying this method to an updated version of the application case in \cite{MMZ2021} concerning the reliability of wind turbines. Furthermore, we are looking at potential active learning methodologies that search for training points which are likely to improve the estimation accuracy of the goal-oriented active subspace matrices.
\par It is worth noting that to control the GP approximation error, we must ensure sufficient exploration of the space \cite{Teckentrup2020}.  In practice, we should add a criterion for minimizing the global GP variance in the high-dimensional space or ensure that the seqOK-AS part does not introduce bias into the final estimation. For the latter point, we could use the active learning criterion developed in SAAS \cite{Wycoff2021} to enrich the design of experiments and potentially couple it with an exploration criterion based on variance reduction. 
The omission of an active enrichment strategy can be justified by the fact that during our sequential strategy, the initial training data, whose size is chosen relative to the high dimension, are effectively used in spaces of progressively reduced dimension. Consequently, the sample size relative to the dimension grows, which somewhat ensures a more accurate GP and thus a more precise dimensionality reduction matrix. However, if the very first GP model(s) are of poor quality, this can introduce a bias in subsequent estimations. 
A design of experiments of sufficient size is necessary to minimize this bias. It should be noted that designs with good projection properties to lower dimensions by construction are recommended, which is feasible if the dimensionality reduction is performed by selecting a subset of the initial dimensions but becomes, to our knowledge, an open problem in the general case.
\par In the same vein, we could also integrate an approach to reduce potential bias arising from uncertainty in the dimensionality reduction and from the fact that the various importance samples, \(\pi^{\varepsilon_j,r_j}(\bs{x})\propto\tilde{L}^{\varepsilon_j}(\mathbf{\hat{W}_{r_j}}^T\bs{x})q(\bs{x})\), which progressively approach the limit state $\{f(\mathbf{x})\leq 0\}$, scarcely explore the full space in regions of low probability for \(q\) and in discarded dimensions. This, in a way, implies that exploration occurs essentially in lower-dimensional spaces. Once dimensionality reduction is applied, very little exploration will be done in the discarded dimensions. In other words, a strategy to better explore the unselected dimensions in the context of reliability analysis would be beneficial.
\par \ValentinCorrections{For problems with a unique failure mode and assuming any error from the seqOK-AS part is negligible, the method we have proposed is efficient for a monotonic failure function or for a convex failure set in the sense that the unique failure mode is always discovered, the Gaussian IS biasing density is relevant thanks to the unicity, and therefore we do not underestimate the failure probability.}
A last perspective concerns cases with multiple failure modes. To extend the approach to cases with several non-negligible failure modes, we could replace the Gaussian approximation of the importance density, which risks leading to an underestimation of the probability, with a Gaussian Mixture Model approximation \cite{kurtz_cross-entropy-based_2013}.
\ValentinCorrections{For a practitioner facing the challenge of multiple failure modes, another option would be to avoid the dimension reduction step, and directly use a biasing density which is more suitable for high-dimensional problems compared to a mixture of Gaussians, such as the von Mises-Fisher-Nakagami mixture \cite{papaioannou2019improved}.
We could also use the distribution induced by the GP approximation (thus avoiding any costly simulations) with a suitable sampling technique. In particular, \cite{MMZ2021} uses a MCMC sampler with a GP approximation of $f(\bs{x})$, coupled with KDR or gKDR dimension reduction; the gradient-free dimension reduction step focuses on the multiple failure modes with the help of a multi-FORM search \cite{der_kiureghian_multiple_1998}.}

\appendix
\section{Performance analysis of seqOK-AS for increasing training size}\label{app:increasing_training_size_regression_benchmark}
\ValentinCorrections{This section compares the performance of our proposed method seqOK-AS versus the baseline approaches OK\_{Py}, OK-AS, and OK-AS+T, which were introduced in Sec. \ref{sec:surrogate_comparison_as}. The comparison is performed for increasing training set size, using the same measures of quality and tested functions described in Sec. \ref{sec:surrogate_validation_numerical_tests}. 
There are 10 repetitions per training size, and the mean and standard deviation are reported for each measure of quality.}
\par
\begin{figure}[htbp]
  \centering
  \begin{tabular}{ccc}
    \begin{subfigure}{0.33\textwidth}
      \includegraphics[width=\linewidth, height=4.7cm]{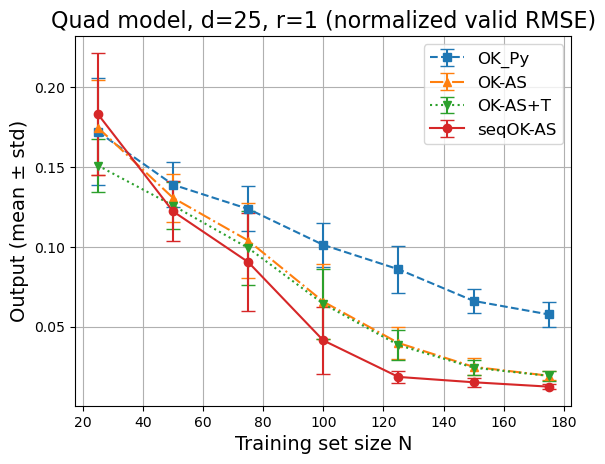}
    \end{subfigure} &
    \begin{subfigure}{0.33\textwidth}
      \includegraphics[width=\linewidth, height=4.7cm]{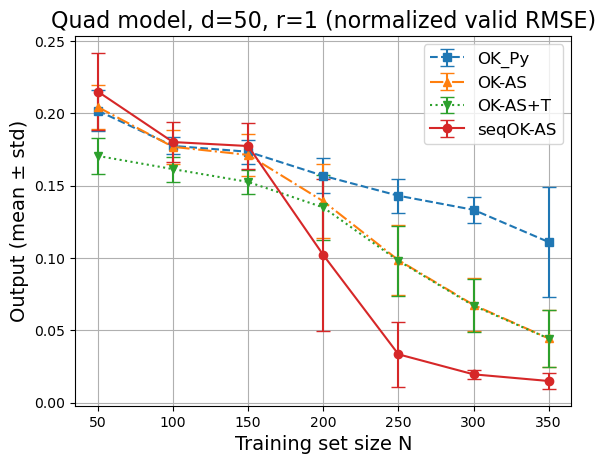}
    \end{subfigure} &
    \begin{subfigure}{0.33\textwidth}
      \includegraphics[width=\linewidth, height=4.7cm]{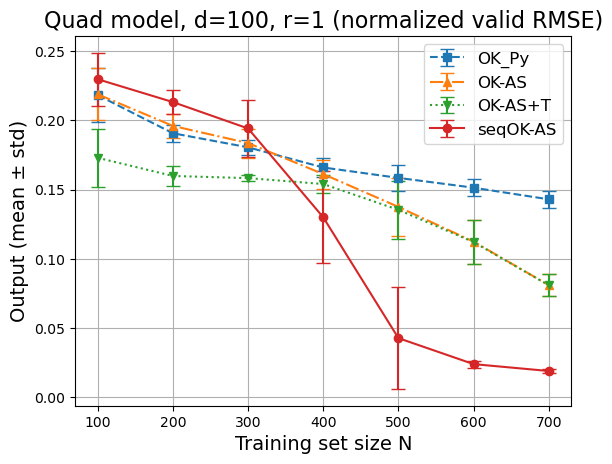}
    \end{subfigure} \\

    \begin{subfigure}{0.33\textwidth}
      \includegraphics[width=\linewidth, height=4.7cm]{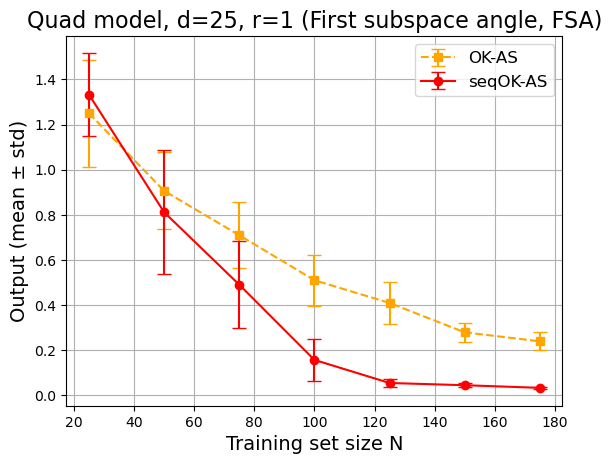}
    \end{subfigure} &
    \begin{subfigure}{0.33\textwidth}
      \includegraphics[width=\linewidth, height=4.7cm]{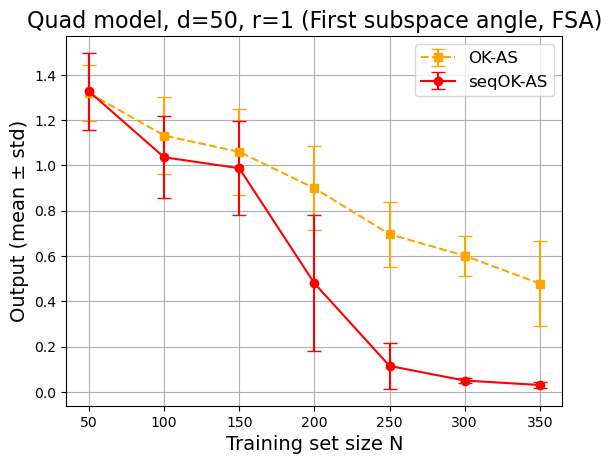}
    \end{subfigure} &
    \begin{subfigure}{0.33\textwidth}
      \includegraphics[width=\linewidth, height=4.7cm]{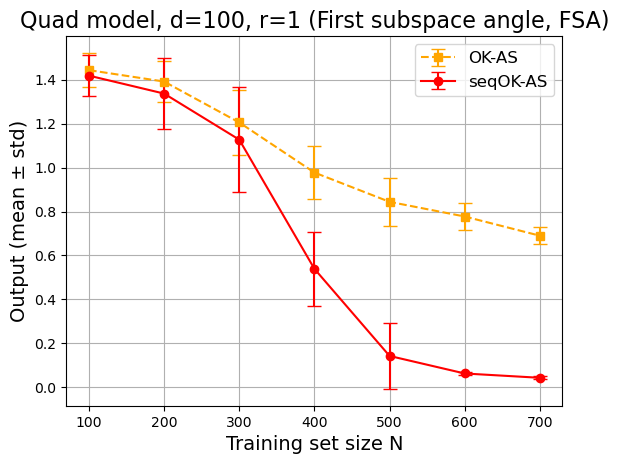}
    \end{subfigure} \\

    \begin{subfigure}{0.33\textwidth}
      \includegraphics[width=\linewidth, height=4.7cm]{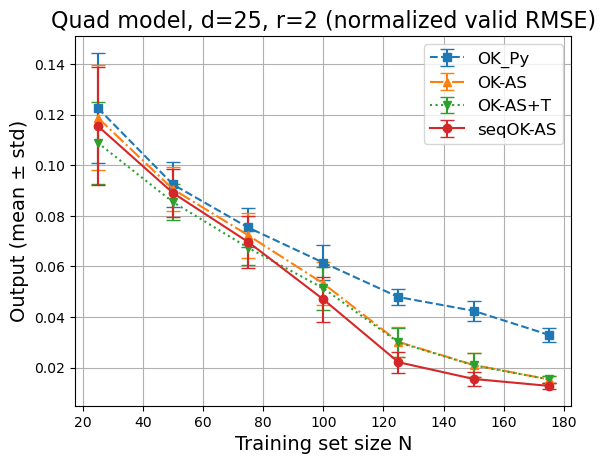}
    \end{subfigure} &
    \begin{subfigure}{0.33\textwidth}
      \includegraphics[width=\linewidth, height=4.7cm]{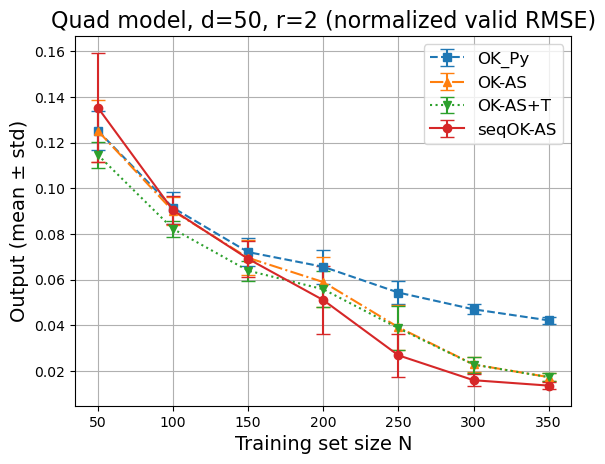}
    \end{subfigure} &
    \begin{subfigure}{0.33\textwidth}
      \includegraphics[width=\linewidth, height=4.7cm]{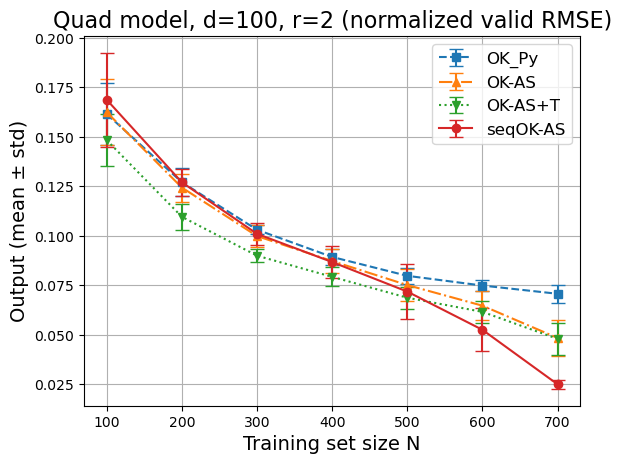}
    \end{subfigure} \\

    \begin{subfigure}{0.33\textwidth}
      \includegraphics[width=\linewidth, height=4.7cm]{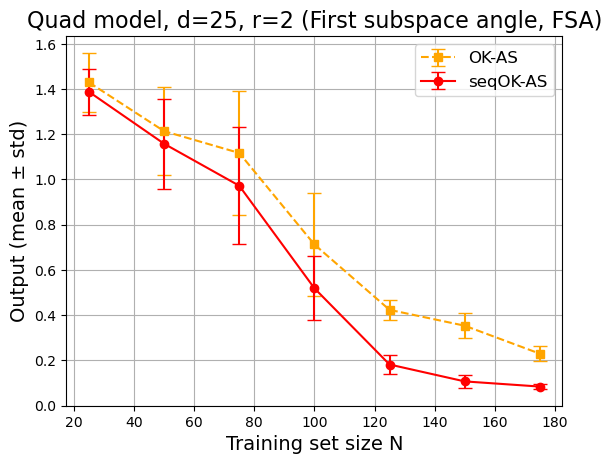}
    \end{subfigure} &
    \begin{subfigure}{0.33\textwidth}
      \includegraphics[width=\linewidth, height=4.7cm]{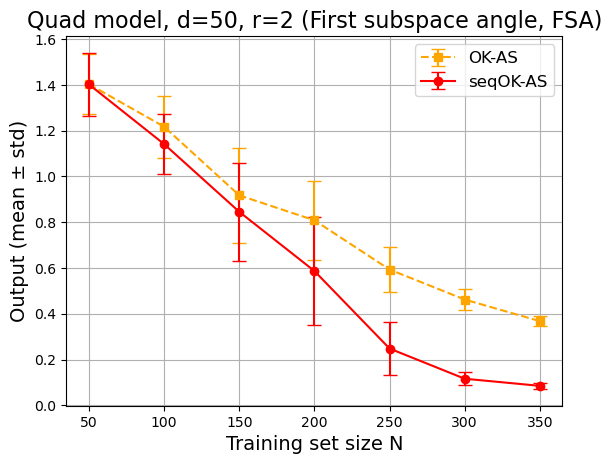}
    \end{subfigure} &
    \begin{subfigure}{0.33\textwidth}
      \includegraphics[width=\linewidth, height=4.7cm]{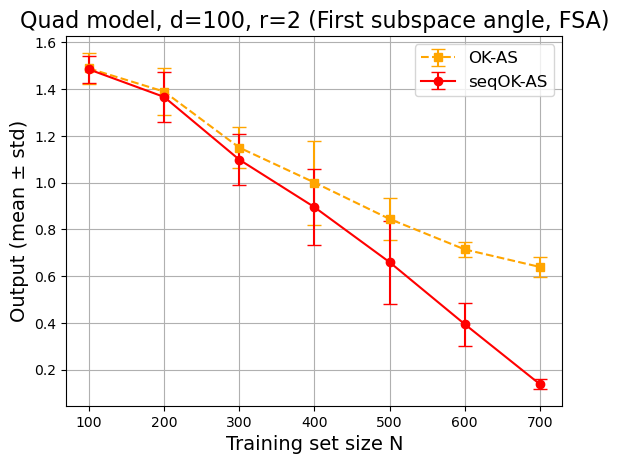}
    \end{subfigure}
  \end{tabular}
  \caption{Method comparison for GPR and AS estimation across different training set sizes for the quadratic test problems}
  \label{fig:rmse_fsa_quadratic_comparison_increasing_training_size}
\end{figure}

\begin{figure}[htbp]
  \centering
  \begin{tabular}{ccc}
    \begin{subfigure}{0.36\textwidth}
      \includegraphics[width=\linewidth, height=4.7cm]{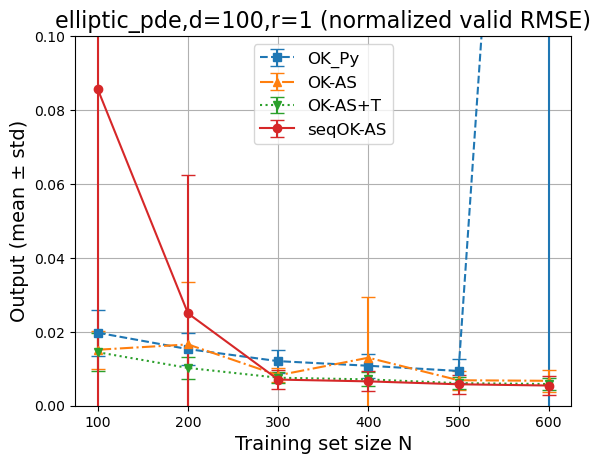}
    \end{subfigure} &
    \begin{subfigure}{0.36\textwidth}
      \includegraphics[width=\linewidth, height=4.7cm]{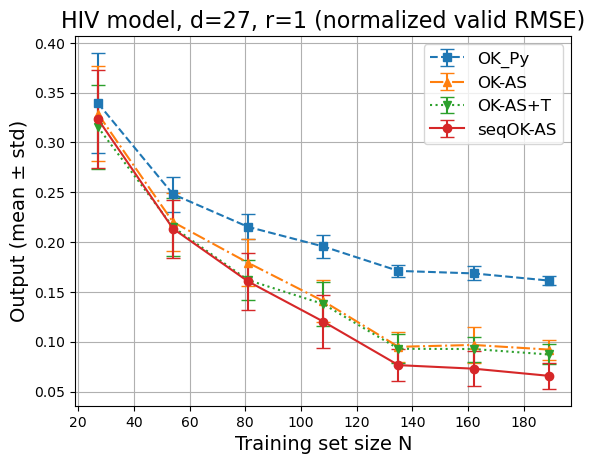}
    \end{subfigure} \\

    \begin{subfigure}{0.36\textwidth}
      \includegraphics[width=\linewidth, height=4.7cm]{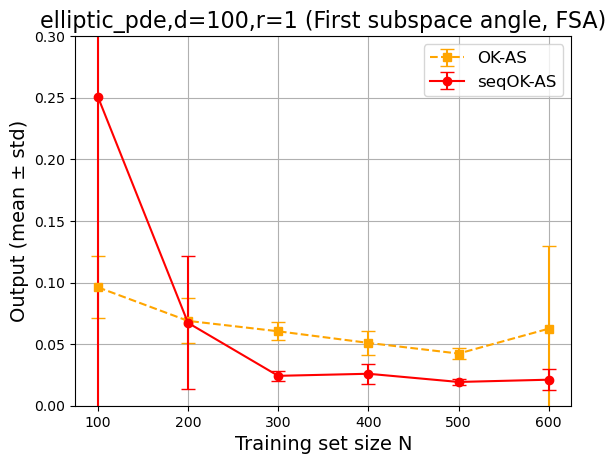}
    \end{subfigure} &
    \begin{subfigure}{0.36\textwidth}
      \includegraphics[width=\linewidth, height=4.7cm]{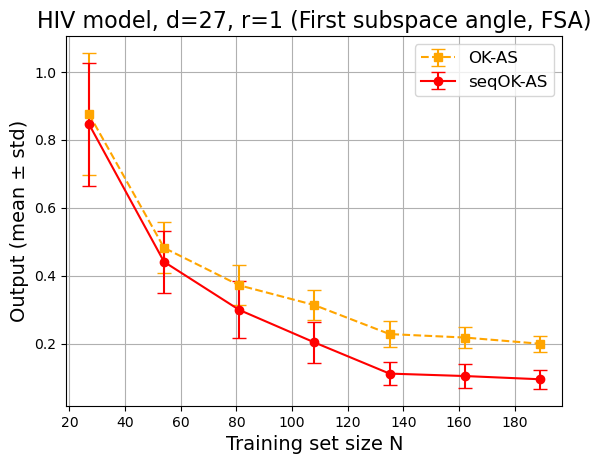}
    \end{subfigure} \\

    \begin{subfigure}{0.36\textwidth}
      \includegraphics[width=\linewidth, height=4.7cm]{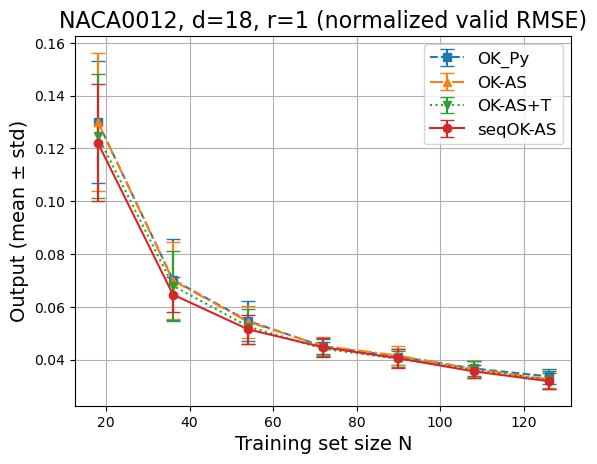}
    \end{subfigure} &
    \begin{subfigure}{0.36\textwidth}
      \includegraphics[width=\linewidth, height=4.7cm]{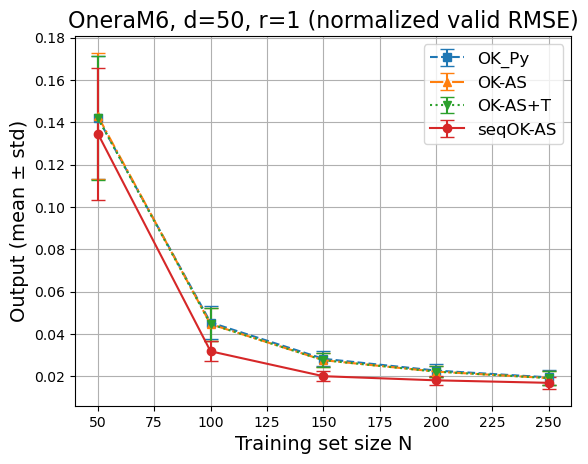}
    \end{subfigure} \\

    \begin{subfigure}{0.36\textwidth}
      \includegraphics[width=\linewidth, height=4.7cm]{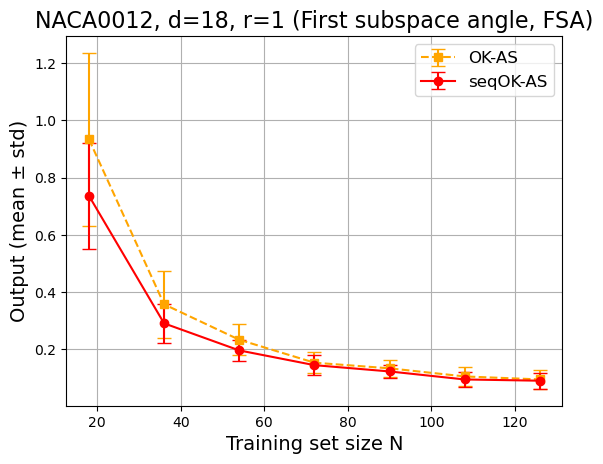}
    \end{subfigure} &
    \begin{subfigure}{0.36\textwidth}
      \includegraphics[width=\linewidth, height=4.7cm]{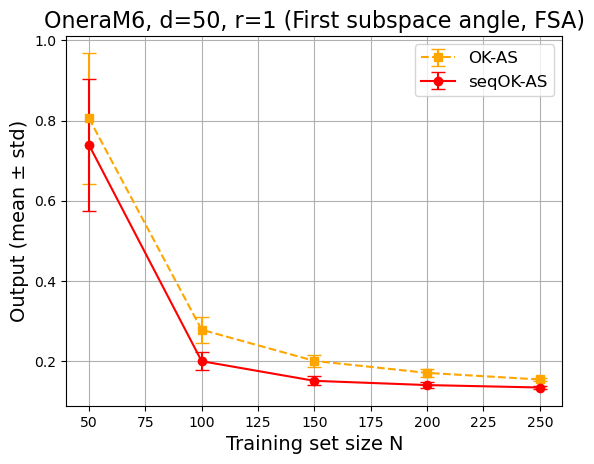}
    \end{subfigure}
  \end{tabular}
  \caption{Method comparison for GPR and AS estimation across different training set sizes for the more realistic problems}
  \label{fig:rmse_fsa_real_world_comparison_increasing_training_size}
\end{figure}

 \ValentinCorrections{From Fig. \ref{fig:rmse_fsa_quadratic_comparison_increasing_training_size}, we draw the following conclusions on the quadratic test problem. First, the performance of seqOK-AS is stable as long as the training size $n$ is large enough. Second, we can see that for large enough $n$, seqOK-AS consistently outperforms the competing methods OK\_{Py}, OK-AS, and OK-AS+T in terms of the average validation RMSE metric over the 10 repetitions. Third, seqOK-AS clearly outperforms OK-AS in terms of active subspace estimation accuracy for large enough $n$, according to the FSA metric. Figure \ref{fig:rmse_fsa_real_world_comparison_increasing_training_size} further confirms these observations on the more realistic test cases. Notably, for the HIV long-term model at a training size of $N=7d=189$, seqOK-AS outperforms all competing methods in terms of both RMSE and FSA.}
\section{Proof of Proposition \ref{prop:alternative_Pf_estimation}}\label{app:proposition_smoothing_dimension_reduction_balance}
\begin{proof}
\ValentinCorrections{First, let us recall the definition of the Hellinger distance
$$
 D_H(\pi,\rho)^2 = \frac{1}{2}\int \left( \sqrt{\pi(\mathbf{x})}-\sqrt{\rho(\mathbf{x})} \right)^2 d \mathbf{x} ,
$$
for any \textbf{unnormalized} densities $\pi,\rho$. We can write
{\allowdisplaybreaks
\begin{align*}
D_H(\pi,\rho)^2
&= \frac{\int d\pi + \int d\rho}{2}  - \int \sqrt{\pi(\mathbf{x})\rho(\mathbf{x})}d \mathbf{x} \\
&= \frac{\left(\sqrt{\int d\pi} - \sqrt{\int d\rho}\right)^2 }{2} + \sqrt{\int d\pi\int d\rho}  - \int \sqrt{\pi(\mathbf{x})\rho(\mathbf{x})}d \mathbf{x} \\
&= \frac{\left(\sqrt{\int d\pi} - \sqrt{\int d\rho}\right)^2 }{2} + \left(1  - \int \sqrt{\frac{\pi(\mathbf{x})}{\int d\pi}\frac{\rho(\mathbf{x})}{\int d\rho}}d \mathbf{x}\right) \sqrt{\int d\pi\int d\rho}\\
&= \frac{\left(\sqrt{\int d\pi} - \sqrt{\int d\rho}\right)^2 }{2} + D_H \left( \frac{\pi}{\int d\pi},\frac{\rho}{\int d\rho}\right) \sqrt{\int d\pi\int d\rho} \\
&\geq \frac{\left(\sqrt{\int d\pi} - \sqrt{\int d\rho}\right)^2 }{2}
\end{align*}}}
\ValentinCorrections{In particular for $\pi(\mathbf{x}) = \mathbb{I}(f(\mathbf{x})\leq 0) q(\mathbf{x})$ and $\rho(\mathbf{x}) =  h( \mathbf{W_r}^\top \mathbf{x} ) q(\mathbf{x})$, we have
\begin{align*}
 \left(\sqrt{P_f} - \sqrt{\widetilde P_f^\varepsilon}\right)^2
 &\leq 2 D_H(\pi,\rho)^2  \leq 4 D_H(\pi,\pi^\varepsilon)^2 + 4 D_H(\pi^\varepsilon,\rho)^2
\end{align*}
where $\pi^\varepsilon(\mathbf{x}) = h^\varepsilon(-f(\mathbf{x})) q(\mathbf{x})$.
On one hand we have
\begin{align*}
 4 D_H(\pi,\pi^\varepsilon)^2
 &= 2\int \left( \sqrt{\mathbb{I}(f(\mathbf{x})\leq 0)}-\sqrt{h^\varepsilon( -f(\mathbf{x}) )} \right)^2 q(\mathbf{x}) d \mathbf{x} \\
 &= 2\int \left( \sqrt{\mathbb{I}(z\leq 0)}-\sqrt{h^\varepsilon( -z )} \right)^2 f_\sharp q(z) d z \\
 &\overset{\eqref{eq:h_CDF}}{=}2\int \left( \mathbb{I}(z\leq 0)- \Phi\left(\frac{z}{\sqrt{2}\varepsilon}\right) \right)^2 f_\sharp q(z) d z \\
 &= 2\int   \Phi\left(-\frac{|z|}{\sqrt{2}\varepsilon}\right)^2 f_\sharp q(z) d z
\end{align*}
where $\Phi(z)=\frac{1}{\sqrt{2\pi}}\int_{-\infty}^z \exp(-t^2/2)d t$ denotes the cumulative distribution function of the standard normal density. Using the standard Gaussian tail bound $\Phi(-z)\leq\frac{1}{2}\exp(-z^2/2)$ for $z\leq0$, we deduce
\begin{align*}
 4 D_H(\pi,\pi^\varepsilon)^2
 &\leq \frac{1}{2} \int   \exp\left(-\frac{z^2}{2\varepsilon^2}\right) f_\sharp q(z) d z  \\
 &= \frac{1}{2} \int   \exp\left(-\frac{f(\mathbf{x})^2}{2\varepsilon^2}\right)  q(\mathbf{x}) d \mathbf{x}
\end{align*}}

\ValentinCorrections{On the other hand, \cite{li2024principal} ensures there exists a function $h:\R^r\to\R_{+}$ such that
\begin{align*}
 4 D_H(\pi^\varepsilon,\rho)^2
 &= 2 \int \left( \sqrt{h^\varepsilon( -f(\mathbf{x}) ) }-\sqrt{h( \mathbf{W_r}^\top \mathbf{x} )} \right)^2 q(\mathbf{x}) d \mathbf{x} \\
 &= 2 P_f^\varepsilon \int \left( \sqrt{\frac{h^\varepsilon( -f(\mathbf{x}) )}{P_f^\varepsilon} }-\sqrt{\frac{ h( \mathbf{W_r}^\top \mathbf{x} ) }{P_f^\varepsilon} } \right)^2 q(\mathbf{x}) d \mathbf{x} \\
 &= 4 P_f^\varepsilon D_H \left(\frac{\pi^\varepsilon}{P_f^\varepsilon},\frac{\rho}{P_f^\varepsilon}\right)^2 \\
 &\overset{\eqref{eq:CDR_hellinger}}{\leq} P_f^\varepsilon C_\mathrm{PI}(q)  \left( \tr(\mathbf{H^\varepsilon}) - \tr( \mathbf{W_r}^\top\mathbf{H^\varepsilon}\mathbf{W_r} ) \right)
\end{align*}
where we recall that $\mathbf{H^\varepsilon}$ is the Fisher information of the (normalized) density $\pi^\varepsilon /P_f^\varepsilon$ relatively to $q$. This yields \eqref{eq:proposition_smoothing_dimension_reduction_balance} and concludes the proof.}
\end{proof}

\clearpage
\section*{Data availability statement}
The authors will share the data and code used in this paper on request.
\bibliographystyle{plain}
\bibliography{bib}

\end{document}